\documentclass[%
 aip,
 jcp, 
 nofootinbib,
 amsmath,amssymb,
 reprint,
]{revtex4-1}

\usepackage{graphicx}
\usepackage{dcolumn}
\usepackage{xcolor}
\usepackage{adjustbox}
\usepackage{import}
\usepackage{booktabs}
\usepackage{mathtools}
\usepackage{comment}
\usepackage{hyperref}
\usepackage{multirow}
\usepackage{esvect}
\usepackage{placeins}
\usepackage{siunitx}   
\DeclarePairedDelimiter\bra{\langle}{\rvert}
\DeclarePairedDelimiter\ket{\lvert}{\rangle}
\DeclarePairedDelimiterX\braket[2]{\langle}{\rangle}{#1\,\delimsize\vert\,\mathopen{}#2}
\usepackage{bm}

\usepackage[utf8]{inputenc}
\usepackage[T1]{fontenc}
\usepackage{mathptmx}
\usepackage{etoolbox}

\usepackage{amsmath}

\DeclareUnicodeCharacter{2500}{\textemdash} 
\DeclareUnicodeCharacter{2212}{-}

\makeatletter
\def\@email#1#2{%
 \endgroup
 \patchcmd{\titleblock@produce}
  {\frontmatter@RRAPformat}
  {\frontmatter@RRAPformat{\produce@RRAP{*#1\href{mailto:#2}{#2}}}\frontmatter@RRAPformat}
  {}{}
}%
\makeatother
\begin{document}

\preprint{AIP/123-QED}

\title
{Aufbau suppressed coupled cluster as a post-linear-response method}

\author{Trine Kay Quady} 
 \affiliation{ 
Department of Chemistry, University of California, Berkeley, California 94720, USA
}
\author{Harrison Tuckman}
 \affiliation{ 
Department of Chemistry, University of California, Berkeley, California 94720, USA
}

\author{Eric Neuscamman}
 \email{eneuscamman@berkeley.edu}

  \affiliation{ 
Department of Chemistry, University of California, Berkeley, California 94720, USA
}
\affiliation{%
Chemical Sciences Division, Lawrence Berkeley National Laboratory, Berkeley, California 94720, USA
}%

\date{\today}

\begin{abstract}
We investigate the ability of Aufbau suppressed coupled cluster
theory to act as a post-linear-response correction to widely used
linear response methods for electronically excited states.
We find that the theory is highly resilient to shortcomings
in the underlying linear response method, with final results
from less accurate starting points nearly as good as those
from the best starting points.
This pattern is especially stark in charge transfer states,
where the approach converts starting points with multi-eV
errors into post-linear-response results with errors on
the order of 0.1 eV.
These findings highlight the ability of Aufbau suppressed
coupled cluster to perform its own orbital relaxations
and raise the question of whether initializing it with an
orbital relaxed reference is worth the trouble.
\end{abstract}

\maketitle

\section{Introduction}
\label{sec:Intro}

Capturing post-excitation relaxations in a molecule's
molecular orbitals (MOs) is often critical for
making accurate excited state predictions.
Indeed, recent examples from configuration interaction (CI),
\cite{shea_communication_2018,shea_generalized_2020,hardikar_self-consistent_2020}
density functional theory (DFT), 
\cite{hait2020excited,hait2020highly,hait2021orbital}
and multi-reference methods
\cite{tran2019tracking,tran2020improving,tran2023exploring,garner2024spin}
provide clear demonstrations of how excited-state-specific MO
optimizations can improve on the partial MO relaxations achieved
through linear response methods.
While these results are encouraging, they are also
somewhat frustrating, because the convenience of getting
many states out of a single linear-response-style diagonalization
gets replaced by state-by-state nonlinear MO optimizations.
Noting that ground state coupled cluster (CC) theory
\cite{crawford2007introduction,bartlett2007coupled,shavitt2009many}
can be remarkably insensitive to the initial choice of MO basis,
\cite{hampel1992comparison,lee1992comparison,bertels2021polishing}
one might wonder whether something similar is true
for the  Aufbau suppressed CC (ASCC) approach
\cite{tuckman_aufbau_2024,
tuckman_improving_2025,tuckman2025NestingArXiv}
to excited states.
After all, both methods can perform their own orbital relaxations
via the CC singles operator,\cite{thouless1960stability}
and so it may be that ASCC, if
initiated from the results of widely used linear response
methods, could combine linear response's ease of use
with the accuracy of state-specific MO relaxations.
As our results demonstrate, ASCC is indeed quite insensitive
to its starting point, allowing it to be used as a high-accuracy,
post-linear-response refinement method in conjunction
with CI singles (CIS), \cite{dreuw2005single}
time-dependent DFT (TD-DFT),
\cite{runge1984density,burke2005time,casida2012progress}
or equation of motion CC (EOM-CC).
\cite{rowe1968equations,stanton1993equation,krylov2008equation}

Excitations that involve substantial orbital relaxations pose
challenges for many widely used excited state methods.
Linear response approaches --- such as 
EOM-CC with singles and doubles (EOM-CCSD), linear response CC,
\cite{monkhorst1977calculation,dalgaard1983some,sekino1984linear,koch1990coupled,rico1993single,koch1994calculation,sneskov2012excited}
and TD-DFT ---
are not able to fully account for these
inherently nonlinear relaxations.
\cite{subotnik2011communication,herbert2023density}
As a result, these methods display reduced accuracy
for charge transfer (CT)
\cite{sobolewski2003ab,dreuw2003long,dreuw2004failure,
mester2022charge,kozma2020new,izsak2020single}
and core excitations.
\cite{coriani2012coupled,frati2019coupled,hait2020highly}
Similarly, the widely used state averaging approach in
complete active space self consistent field (CASSCF) theory ---
in which all states are forced to share the same MO basis ---
creates significant errors for CT states,
\cite{domingo2012metal,pineda2019excited,tran2019tracking,tran2020improving}
which require different orbital polarizations than the
ground state and most other excited states.
These difficulties are not just an issue for excitation
energies: they can also produce large errors in the
excited state potential energy surfaces
\cite{tran2020improving,clune2025emlc}
that control photochemical reactions.

Instead of relying on ground state or partially relaxed
orbitals, an increasing number of approaches seek to
fully relax the MOs in the presence of the excitation.
These approaches include a number of DFT,
\cite{hait2020excited,
hait2020highly,
hait2021orbital,
ye2017sigma,
ye2019half,
bagus1965scf,
Pitzer1976scf,
argen1991xray,
besley2009self,Shaik1999,
kowalczyk2013excitation,
kowalczyk2011assessment,
zhao2019density,
levi2020variational,
kempfer2022role,
gilbert2008self,
barca2018simple,
carter2020state}
variational Monte Carlo,
\cite{zhao2016efficient,robinson2017excitation,blunt2017charge,shea2017size,
pineda2019excited,garner2020variational,otis2020hybrid,shepard2022double,
otis2023optimization,otis2023promising,pathak2021excited,entwistle2023electronic}
CASSCF,
\cite{tran2019tracking,
tran2020improving,
tran2023exploring,
hanscam2022applying,
roos_towards_1992,
boyn_elucidating_2022}
perturbation theory,
\cite{clune2020n5,
clune2023studying,
clune2025emlc}
CC,
\cite{mayhall2010multiple,
zheng2019performance,
lee2019excited,
damour2024state,
kossoski2021excited,
tuckman_excited-state-specific_2023}
and CI methods.
\cite{shea_communication_2018,
shea_generalized_2020,
hardikar_self-consistent_2020,
liu2012communication,
kossoski2022state,
kossoski2023seniority,
burton2022energy,
tsuchimochi2024CISthenCIS}
In general, these excited-state-specific methods tend to be
more accurate for CT and core excitations than their linear
response counterparts, but this improved accuracy comes at a cost.
Unlike linear response --- in which a single, well-conditioned diagonalization
yields many excited states at once --- these state-specific methods
must perform individual nonlinear MO optimizations for each
excited state separately.
Although such optimizations have proven effective in many cases,
convergence to the correct root is not trivial, \cite{burton2022energy}
and there are known cases where convergence
has failed. \cite{clune2023studying}
Ideally, it would be possible to combine the best of both worlds by
pairing the ease of use of linear response with the high accuracy
of state-specificity.

Although ASCC theory was initially motivated as a way to
build a CC theory atop excited state mean field (ESMF) theory,
\cite{tuckman_aufbau_2024}
its ability to incorporate its own state-specific
orbital relaxations via its singles operator
raises the possibility of instead pairing it
with starting points from linear response theory.
The basic idea is that the linear response method, via
its convenient diagonalization, can produce a qualitatively
correct starting point that ASCC then improves upon.
Of course, this approach may simply shift the nonlinear
MO optimization challenge into the ASCC equations,
but this difficulty should be mitigated by linear response
methods that at least partially relax the orbitals.
For example, while CIS and TD-DFT are only
able to relax the shapes of the hole and particle orbitals,
EOM-CCSD is able to capture limited relaxation effects for
all orbitals via its CI doubles operator, which would
presumably lessen the difficulty of finalizing these
relaxations with ASCC.
Thus, we hope to use ASCC as a post-linear-response
method that benefits from linear response's ease of use
while delivering improved, state-specific accuracy.

To test this approach, we have paired ASCC with
CIS, TD-DFT, and EOM-CCSD and tested the pairings
on 145 singlet excitations spanning valence,
Rydberg, and CT excited states.
As in our previous studies of ASCC, we benefit
enormously from the high quality QUEST database
\cite{loos_mountaineering_2018,loos_mountaineering_2020}
as well as the CT benchmark of 
Szalay and coworkers.~\cite{kozma2020new}
Remarkably, we find that even when ASCC is
initiated from TD-DFT or CIS starting points
whose CT excitation energy errors are larger than 2 eV,
the final ASCC results are typically on par with
those initiated from ESMF or EOM-CCSD.
Although there does remain a (very) small
accuracy advantage to starting from ESMF,
the results we present below call into question
whether going to the trouble of performing ESMF's
nonlinear orbital optimization is worthwhile
in conjunction with ASCC.
For singly-excited singlet excitations at least,
the data strongly suggest that the most productive
use of ASCC may well be as a post-linear-response method.

\section{Theory}
\label{sec:Theory}

\subsection{Setting up ASCC}
\label{sec:ascc-setup}

Like many CC theories, ASCC is motivated by the strong formal
properties offered by an ansatz in which exponentiated
operators act on a single-determinant reference.
These properties include systematic improvability,
size consistency, size extensivity, and, for excitation energies,
size intensivity.
Unlike most CC approaches, the ASCC wave function includes
both an excitation operator $\hat{T}$ and a de-excitation
operator $\hat{S}^\dagger$.
\begin{align}
    \label{eq:ASCC-psi-simple}
    \ket{\Psi_\text{ASCC}} &= e^{-\hat{S}^\dagger} e^{\hat{T}} \ket{\phi_0}
    = \left( 1 + \hat{T} -\hat{S}^\dagger \hat{T}
                + ... \right) \ket{\phi_0} 
\end{align}
The key idea is that, when acted on the Aufbau determinant $\ket{\phi_0}$,
the term $\hat{S}^\dagger\hat{T}$ produces a second copy of that
determinant via an excite-then-de-excite process.
So long as the sign on that copy is negative, it can wholly or partially
cancel the main Aufbau term in the expansion, thus producing a state
in which the Aufbau determinant is absent or present with only a small coefficient.
Like other CC theories, the working equations involve a similarity transformed
Hamiltonian.
\begin{align}
    \label{eq:sim-sim-H}
    \bar{H} &=
    e^{-\hat{T}} e^{\hat{S}^\dagger} \, \hat{H} \, e^{-\hat{S}^\dagger} e^{\hat{T}} \\
\label{eq:E_ASCC}
    E_\text{ASCC} & = \bra{\phi_0}\bar{H}\ket{\phi_0} \\
\label{eq:amp_eqns}
    0 & = \bra{\mu}\bar{H}\ket{\Phi_0}  
\end{align}
Here $\mu$ stands for any of the excited determinants
corresponding to the individual excitation operators placed in $\hat{T}$.
Partially linearized ASCC (PLASCC) uses these
same working equations, but with certain nonlinear
diagrams neglected in a balancing effort motivated by
perturbative analysis.
\cite{tuckman_improving_2025}

To decide what to include in $\hat{T}$, we follow perturbative arguments
that first require us to define the zeroth order pieces of the wave function.
In practice, we start with an initial guess for the excited state
(e.g., from CIS, TD-DFT, ESMF, or EOM-CCSD)
and truncate it down to only its most significant
configuration state functions (CSFs).
We then define the zeroth order pieces of $\hat{S}$ and $\hat{T}$ so that,
when truncated to zeroth order, the ASCC wave function matches the truncated
initial guess, which we call our starting point.
In the simplest case, the starting point has just one CSF in which
an electron has been promoted from the hole orbital $h$
to the particle orbital $p$, which gives us a particularly
simple zeroth order setup. \cite{tuckman_aufbau_2024}
\begin{align}
    \label{eq:S0}
    \hat{S} =
    \hat{S}^{(0)} &= \frac{1}{\sqrt{2}}\left(\hat{p}^\dagger_\uparrow \hat{h}_\uparrow + \hat{p}^\dagger_\downarrow \hat{h}_\downarrow \right)  \\
    \label{eq:T0}
    \hat{T}^{(0)} &= \hat{S} - \frac{1}{2}\hat{S}^2
\end{align}
In cases where there are multiple CSFs in the truncated starting point,
the zeroth order setup is more complicated, \cite{tuckman_improving_2025}
but key characteristics are maintained.
Specifically, $\hat{S}=\hat{S}^{(0)}$ will only contain zeroth order pieces,
it will be a one-body operator, and it will only excite to and from
orbitals that are half-occupied in at least one of the starting point CSFs.

With the zeroth order wave function chosen, we then define a suitable
Hamiltonian partitioning \cite{tuckman_improving_2025}
and flesh out $\hat{T}$ by incorporating all excitation operators
whose amplitudes appear at first order in perturbation theory.
In the one-CSF case, this approach leads to including
all singles and doubles excitations, as well as a small subset
of triples (in multi-CSF cases, there are also small subsets
of higher excitations).
So long as there are only $O(1)$ primary CSFs, the approach will
have CCSD's $O(N^6)$ cost scaling and contain only $O(N^3)$
amplitudes beyond the doubles, \cite{tuckman_improving_2025}
which clearly motivates us to seek compact starting points
whose wave functions are dominated by a small number of CSFs.
As we will now discuss, ESMF is far from the only way to get such
starting points, as they are also readily available from a variety of
different linear response theories.

\subsection{Starting from CIS}
\label{sec:CIS-start}

The CIS wave function is defined by its coefficient
matrix $\bm{C}$, which can be indexed by an occupied
index $i$ and virtual index $a$.
\begin{align}
    \label{eqn:cis-wfn}
    \ket{\Psi_\text{CIS}} = \sum_{ia} C_{ia} \ket{\phi_i^a}
\end{align}
Here $\ket{\phi_i^a}$ is the singly-excited singlet CSF
in which an electron has been excited from orbital $i$
to orbital $a$.
As shown by Martin, \cite{martin2003natural} the
natural transition orbitals (NTOs) for CIS are derived
from a singular value decomposition (SVD) of the
matrix $\bm{C}$, as $\bm{C}$ is also the CIS
transition density matrix (TDM).
If we work in this NTO basis, the wave function becomes
\begin{align}
    \label{eqn:cis-nto}
    \ket{\Psi_\text{CIS}} = \sum_{k} \sigma_{k} \ket{\phi_k^{\tilde{k}}},
\end{align}
where $k$ and $\tilde{k}$ label the occupied NTO and
virtual NTO corresponding to singular value $\sigma_k$.
As a truncated SVD is (in the least-squares sense)
the best low-rank approximation of a matrix,
\cite{eckart1936approximation}
we produce an optimal truncation of the CIS
state for our ASCC starting point by
moving into the CIS NTO basis,
dropping small singular values
(we use a threshold of 0.2),
and renormalizing.
 
\subsection{Starting from ESMF}
\label{sec:ESMF-start}

For ESMF, we will again prepare the ASCC starting point
via a truncated SVD approach, but, due to ESMF's orbital
relaxations, the matrix in question is not the ESMF TDM.
We start in the relaxed ESMF orbital basis,
in which the ESMF state takes the same form as
Eq.\ (\ref{eqn:cis-wfn}).
\cite{hardikar_self-consistent_2020}
From there, we follow the same procedure of taking the
SVD of the ESMF $\bm{C}$ matrix and using the singular
vectors to rotate to an MO basis in which the CI
matrix becomes diagonal, as in Eq.\ (\ref{eqn:cis-nto}).
Since the ESMF MOs differ from the ground state Hartree
Fock MOs, $\bm{C}$ is not the ESMF TDM, and so
this new basis resulting from an SVD of $\bm{C}$
is not the NTO basis.
We have therefore instead referred to it as ESMF's
transition orbital pair (TOP) basis when employing it
in other post-ESMF methods. \cite{clune2020n5}
Nonetheless, the ESMF TOP basis is a close cousin to
the CIS NTO basis, and it will play the same role in our
ASCC starting point preparation.
After moving to the TOP basis, we truncate small singular
values and renormalize to produce our starting point.

\subsection{Starting from EOM-CCSD}
\label{sec:EOM-CCSD-start}

The nature of the EOM-CCSD TDM creates complications
for our approach.
To start, there are actually two different TDMs that
appear in the theory,
\cite{stanton1993equation}
owing the fact that the left and right eigenvectors
of its non-Hermitian similarity transformed
Hamiltonian are not the same.
Further, correlation effects prevent either of
these TDMs from having the simple
occupied-by-virtual rectangular
structure of the CIS TDM, which in turn means that
their singular vectors do not define a change
of MO basis that keeps occupieds and virtuals separate.
In a state where the TDM has just one singular value
significantly different than zero, the other singular
vectors can strongly mix occupied and virtual
orbitals with each other, creating an MO basis
that is difficult to interpret and poorly suited
for a theory that intends to define $\hat{T}$ as
being built from operators that, roughly speaking,
excite from the occupied space to the virtual space.

To overcome these challenges, we define our EOM-CCSD-based
ASCC starting point using a combination of the ``right-hand''
$\bra{L_\text{ground}}\cdots\ket{R_\text{excited}}$
EOM-CCSD TDM \cite{stanton1993equation}
and the EOM-CCSD excited state natural orbitals (NOs).
We start by performing an SVD of the TDM and
identifying above-threshold singular values.
We then form a list of (not yet orthogonal)
occupied orbitals that begins with the
above-threshold left hand singular vectors in
descending order, followed by the natural orbitals
with occupations near two.
We form a second list of (not yet orthogonal) virtual
orbitals from the above-threshold right hand singular
vectors, followed by the natural orbitals with
occupations close to zero.
Note that the natural orbitals corresponding to the
retained left and right singular vectors
(those with the largest overlaps) are not used.
Finally, we combine these lists by interleaving them
--- occupied, virtual, occupied, virtual, etc ---
and then performing a Gram–Schmidt process to
arrive at an orthonormal MO basis.
As in Eq.\ (\ref{eqn:cis-nto}), our starting point is
then built from the CSFs corresponding to the
above-threshold singular values.

\subsection{Starting from TD-DFT}
\label{sec:TD-DFT-start}

Although practical implementations of TD-DFT tend to lead
to TDMs with nonzero entries only in the occupied-virtual
and (if not using Tamm-Dancoff) virtual-occupied blocks
\cite{furche2001density}
and so will yield NTOs that do not mix occupieds with
virtuals, exact TD-DFT would match full CI which,
like EOM-CCSD, can mix occupieds and virtuals in its NTOs.
We therefore choose to prepare our truncated TD-DFT
starting point in the same manner as for EOM-CCSD.
Specifically, we have employed the Tamm-Dancoff TDM
and excited state NOs in the same manner as we employed
the EOM-CCSD TDM and excited state NOs above.
In tests, we have verified that this approach leads to
excitation energies that are virtually indistinguishable
(within 0.01 eV) from the alternative option of
preparing the starting point via the CIS approach
with the Tamm-Dancoff TDM swapped in for the CIS TDM.

\begin{figure}[]
    \centering
    \includegraphics[width=0.95\linewidth]{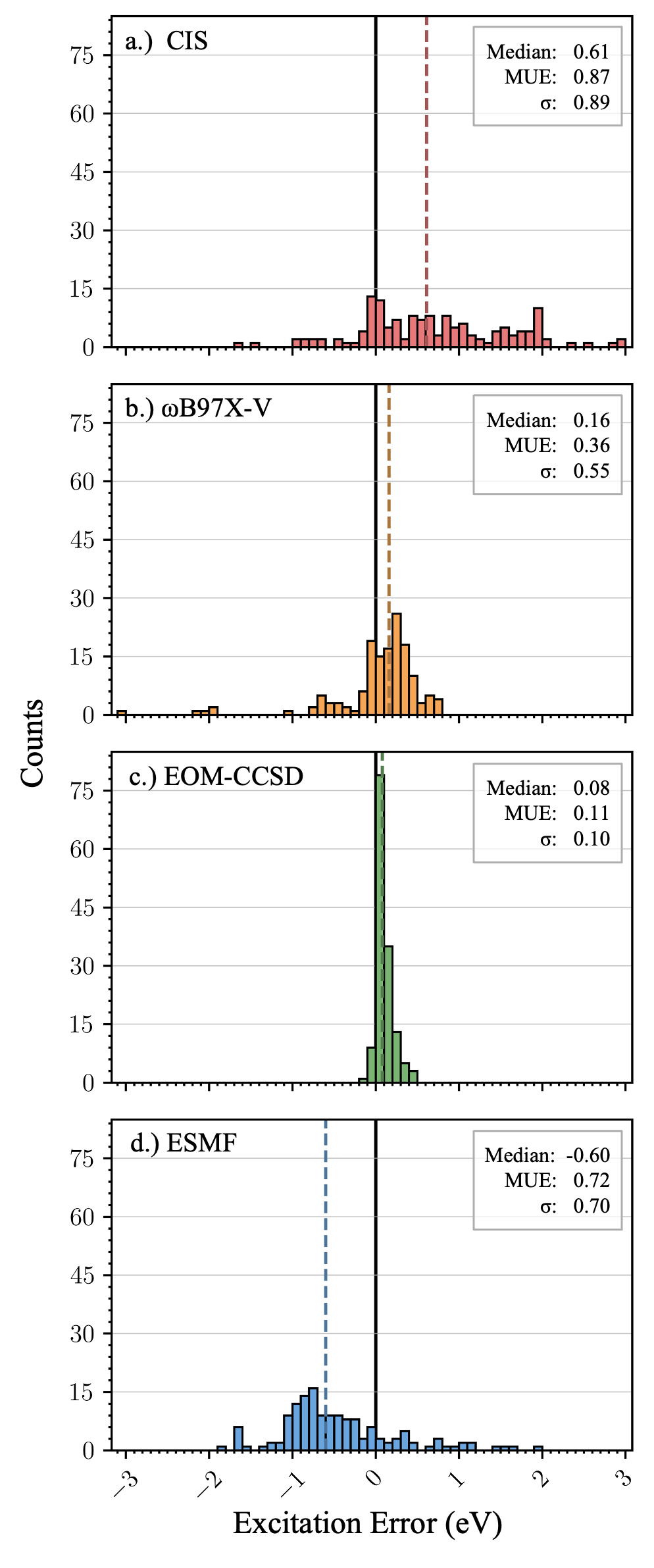}
    \caption{Distributions of excitation energy errors for 
    CIS,  TD-DFT/$\omega$B97X-V,  EOM-CCSD, and  ESMF
    across all states, which include the seven charge transfer states
    and all states in our reduced test set of singlet valence and Rydberg excitations.
    The dashed line shows a method's median error, while the insets report this median,
    the mean unsigned error (MUE),
    and the standard deviation ($\sigma$).
    }
    \label{fig:EE}
\end{figure}

\section{\label{sec:Results}Results}

\subsection{Computational Details}

CIS, ESMF, TD-DFT, and EOM-CCSD vertical excitation energies,
and their  corresponding ASCCSD and PLASCCSD counterparts,
were compared against benchmark results from three datasets containing singly excited valence, Rydberg, and charge transfer states.
CIS calculations were performed with PySCF~\cite{sun2020recent}
while EOM-CCSD and TD-DFT/$\omega$B97X-V \cite{mardirossian2014omegab97x}
calculations were performed using Q-Chem 6.2. \cite{epifanovsky2021software}
The TDMs and NOs used to derive ASCC starting points from EOM-CCSD
and (TDA) TD-DFT were produced by Q-Chem's excited state analysis module.

The benchmark results include the QUEST small and medium molecular
benchmark datasets, \cite{loos_mountaineering_2018,loos_mountaineering_2020}
which together contain 188 single excitations, including singlet
valence and Rydberg excited states,
while the remaining dataset consists of seven charge transfer states.
\cite{tuckman_improving_2025}
In this work and in the benchmarks, valence and Rydberg calculations were performed in the aug-cc-pVDZ basis while charge transfer calculations were performed in the cc-pVDZ basis.
While the benchmark sets used the frozen core approximation and the present calculations did not, freezing the core typically has a 0.02 eV or smaller effect on the excitation energy~\cite{loos_mountaineering_2018,loos_mountaineering_2020}
and so does little to interfere with the comparisons.

In prior work, \cite{tuckman_improving_2025}
the 188 states from the two QUEST sets were filtered down to
144 states by focusing on one- and two-CSF singly excited states,
excluding a handful of cases that lacked ESMF, ASCC, or PLASCC solutions,
and excluding two-CSF states in molecules too large for our pilot two-CSF
implementation to handle.
In the present study, different starting points prove to be better and worse
for different states, and in a number of cases one or more of the new starting
points was poor enough that ASCC or PLASCC failed to converge, even though
they had converged when starting from ESMF.
Some starting points also proved to have more than two CSFs, which is beyond
the capabilities of our current ASCC code base.
In order to focus on cases where a comparison across at least three starting
points was possible, we have therefore also excluded states where two or more
of the starting points failed to converge or had more than two large CSFs.
This winnowing leaves us considering a reduced test set of
138 valence and Rydberg excitations (see SI for details).
In addition, we consider all seven charge transfer states from and use the
the same convergence criteria and two-ansatz averaging procedure 
from the recent perturbative analysis study. \cite{tuckman_improving_2025}

\subsection{Starting method energies}
\label{sec:start-energies}

To appreciate the degree to which ASCC and PLASCC are able to offer
improvement over different starting points, we first analyze the
accuracy of the starting point methods themselves.
As seen in Figure \ref{fig:EE}, EOM-CCSD is by far the most
accurate, with TD-DFT/$\omega$B97X-V showing intermediate
accuracy.
Both CIS and ESMF show relatively poor accuracy, but in interestingly
different ways.
Lacking orbital relaxations, CIS tends to error high.
\cite{dreuw2005single}
In contrast, ESMF tends to error low, which can be explained by the
fact that, while it lacks most correlation effects, it does capture
the strong correlation between the two electrons involved in the
excitation's open-shell singlet.
The ground state energy from Hartree Fock that is used to get the
ESMF excitation energy difference lacks all correlation effects, and
so we would expect ESMF to on average error low by about one electron
pair's worth of correlation, which is roughly what we see.

Although they are in some cases hidden in the distribution, the
charge transfer states prove especially challenging for
the four starting methods.
In CIS and EOM-CCSD, they produce errors at the high end of the
error distribution, whereas in TD-DFT they produce large negative
errors that stand out clearly on the left hand side of its
error distribution.
Given EOM-CCSD's especially high accuracy for valence and Rydberg
states, we would expect the largest improvements from subsequent
ASCC and PLASCC improvements to occur for the charge transfer states.
Based on the accuracy reached in previous studies,
we would hope that ASCC and especially PLASCC could make
significant improvements on the other three starting points
in all three categories of states: valence, Rydberg, and charge transfer.
We will begin with the first two categories, after which we will
turn our attention to the especially challenging case of charge transfer.

\begin{figure*}[]
    \centering
    \includegraphics[width=0.95\linewidth]{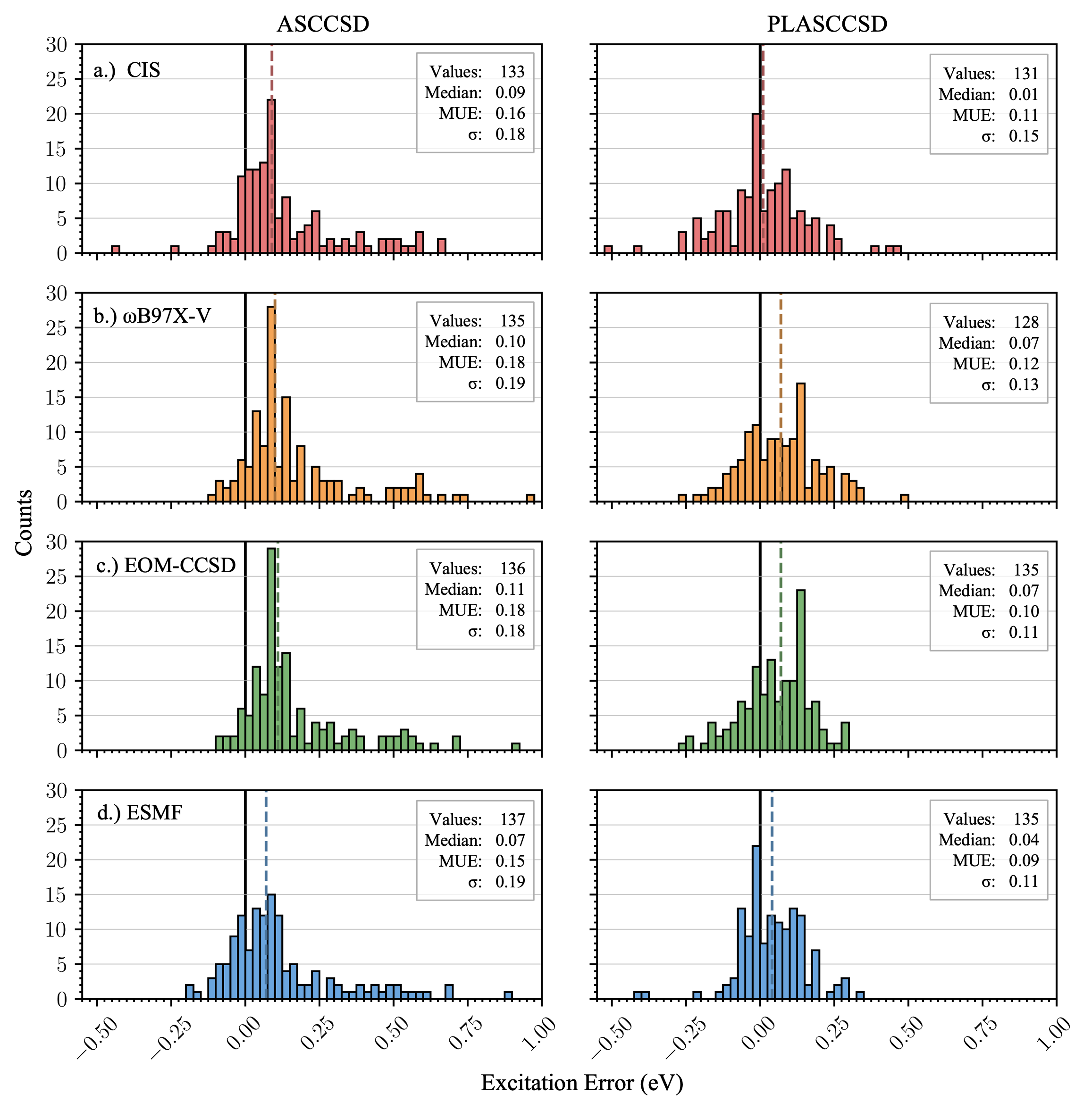}
    \caption{Excitation energy error distributions for ASCCSD (left) and PLASCCSD (right) on our reduced test set of valence and Rydberg excitations. 
    From top to bottom, the starting points used were
    CIS, $\omega$B97X-V, EOM-CCSD, and ESMF. 
    The dashed lines show the median errors, while the insets report these,
    the mean unsigned error (MUE), the standard deviation ($\sigma$),
    and the number of states in which that pairing of methods converged.
    }
    \label{fig:smallMedQUEST_all}
\end{figure*}

\subsection{Valence and Rydberg states}
\label{sec:ResultsQUEST}

\begin{figure*}[]
    \centering
    \includegraphics[width=0.95\linewidth]{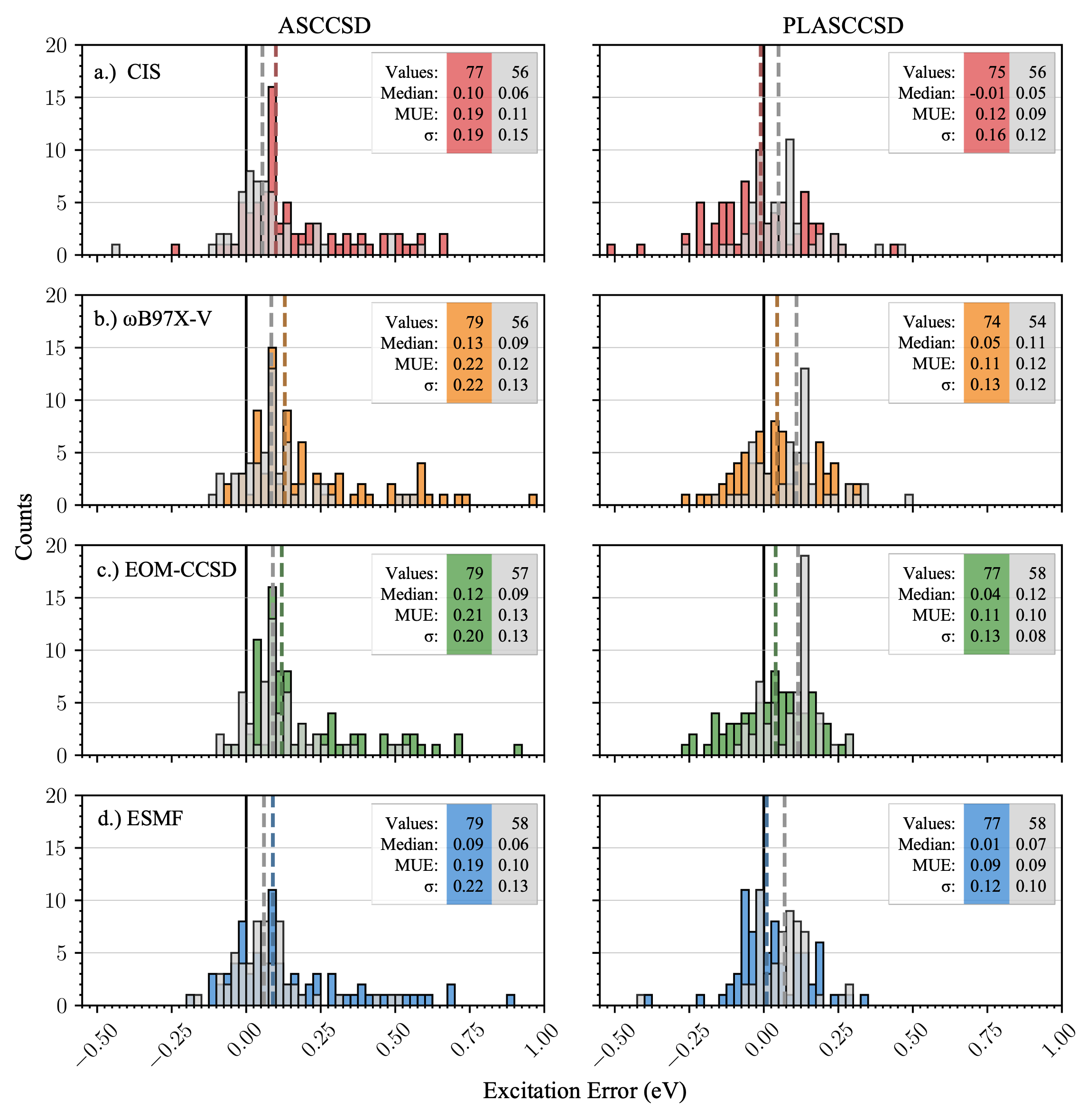}
    \caption{ASCC (left) and PLASCC (right) excitation energy error distributions
    for different starting points broken out by state type:
    results for valence states are shown in color
    and those for Rydberg states in gray.
    Dashed lines show median errors, while these along with the MUEs,
    standard deviations, and number of converged states
    are shown in the insets.
    }
    \label{fig:small_med_rydberg}
\end{figure*}

Remarkably, when we apply ASCC and PLASCC to the ESMF, CIS, TD-DFT, and EOM-CCSD
starting points, the resulting accuracies are quite similar and amount
to a significant improvement over three of the four starting methods.
As seen in Figure \ref{fig:smallMedQUEST_all}, the median error and
mean unsigned error (MUE) for ASCC is quite insensitive to the starting point used.
A similar result occurs for PLASCC, which also proves again to be more
accurate than ASCC.
In PLASCC, we also see a small accuracy advantage when starting from EOM-CCSD and
especially ESMF, which we presume is thanks to the fact that
these starting points are better at building in orbital relaxation effects
than CIS or TD-DFT.
There also appears to be a small advantage in terms of how easy it is to
converge our CC equations, with EOM-CCSD and ESMF appearing to offer
starting points closer to the ASCC or PLASCC solution points, or, at least, starting
points from which the ASCC and PLASCC calculations are more likely to converge.

Overall, however, the main takeaway is not in the differences
between results from different starting points, but in how similar they are.
PLASCC, for example, converts starting points from methods with MUEs of
0.87, 0.72, 0.36, and 0.11 eV in Figure \ref{fig:EE} into results with
MUEs of 0.12 eV or less.
These uniformly small errors come despite the fact that the starting points
contain very different amounts of post-excitation orbital relaxations:
CIS and TD-DFT only relax the shape of the particle and hole orbital,
EOM-CCSD also contains linearized relaxations for other orbitals, and
ESMF offers full nonlinear relaxations for all orbitals.
Nonetheless, Figure \ref{fig:smallMedQUEST_all} shows that the ASCC and PLASCC
error distributions are nearly independent of starting point, which appears
to confirm that their exponentiated singles operators are indeed capable
of building in orbital relaxations that may be missing in the starting point.
We find it particularly noteworthy that when starting from CIS and TD-DFT,
the simplest and most widely available starting points, PLASCC cleans things
up to the point that typical errors are reduced to one or two tenths of
an eV, and are nearly as good as when starting from EOM-CCSD and ESMF.

\begin{figure*}[]
    \centering
    \includegraphics[width=0.95\linewidth]{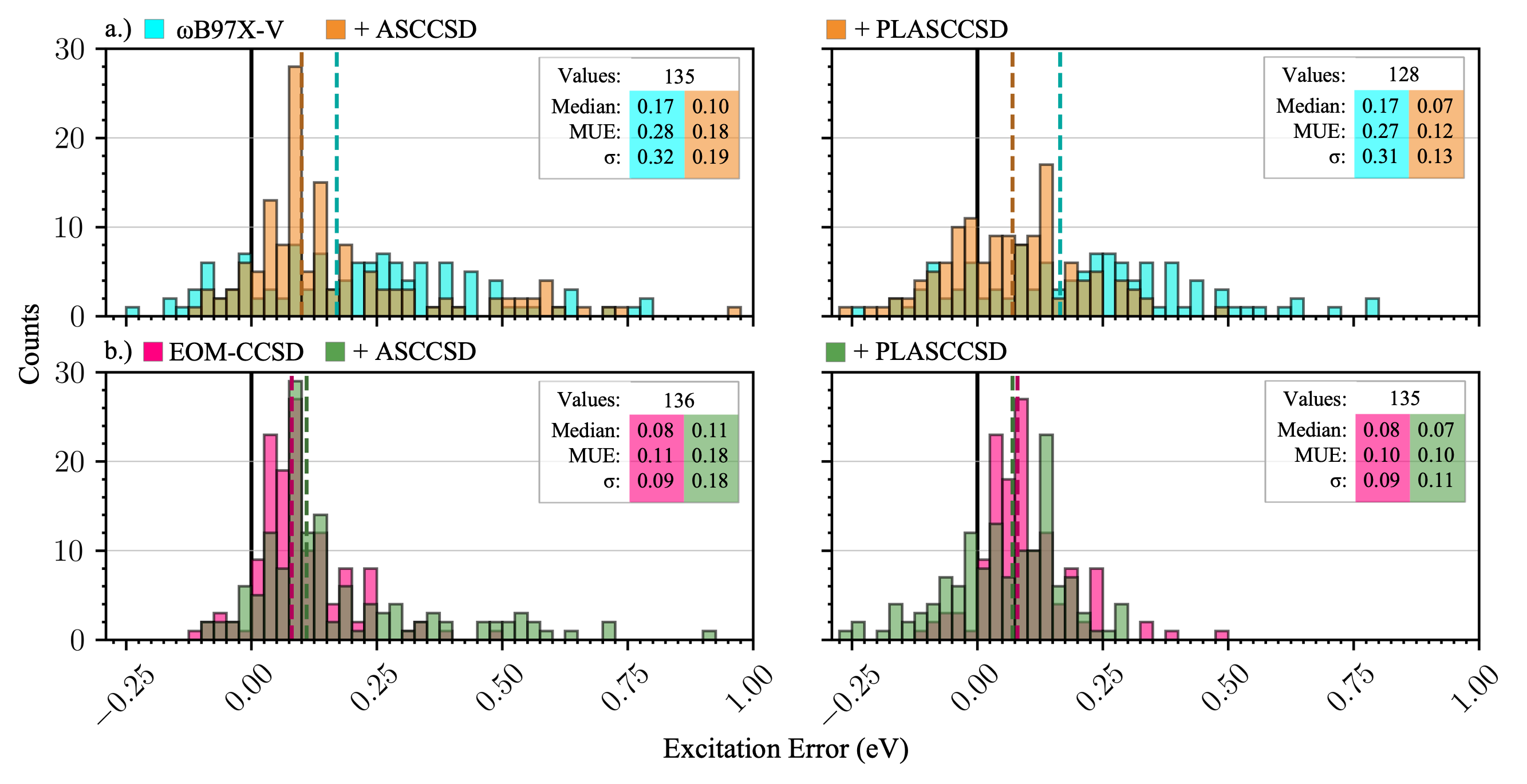}
    \caption{Excitation energy error distributions on our reduced valence
    and Rydberg set for TD-DFT/$\omega$B97X-V (a) and  EOM-CCSD (b)
    before and after post-linear-response treatments with ASCC and PLASCC.
    Dashed lines show median errors, with additional statistics in the insets.
    }
    \label{fig:EOM-wB97-EEvsPL-ASCC}
\end{figure*}

In Figure \ref{fig:small_med_rydberg}, we break down these results into
separate histograms for valence and Rydberg states.
Even within each category, we see the same pattern:
the results are quite insensitive to the choice of starting point.
This breakout also reveals that PLASCC's accuracy advantage over
ASCC appears to be concentrated in valence states, as the two
approaches show much more comparable errors in Rydberg states.
Again, PLASCC shows the best accuracy overall, and its accuracy
is slightly higher for the EOM-CCSD and ESMF starting points, but
not by much.
It is especially noteworthy to see that, in contrast to TD-DFT's
well known difficulties in Rydberg states, \cite{laurent2013td}
PLASCC based on a TD-DFT starting point performs almost as
well for these states as it does when based on ESMF.

Finally, in Figure \ref{fig:EOM-wB97-EEvsPL-ASCC}, we show
the combined valence and Rydberg error distributions for
TD-DFT and EOM-CCSD before and after applying our
ASCC or PLASCC post-linear-response treatment.
While ASCC improves on the TD-DFT starting point, it
goes in the wrong direction when starting from the
already high-accuracy EOM-CCSD results.
PLASCC, on the other hand, more or less maintains the
high accuracy of EOM-CCSD while significantly improving
on TD-DFT.
At first glance, it therefore appears that there may
not be much reason to favor PLASCC over EOM-CCSD for
valence and Rydberg states.
However, given the recent findings that nested
PLASCC offers similarly high accuracies at greatly
reduced cost, \cite{tuckman2025NestingArXiv}
it may in future be worthwhile to apply the nested
theory as a low-cost, post-linear-response correction
atop TD-DFT.

\subsection{Charge transfer states}
\label{sec:ResultsCT}

\begin{figure}[h!]
    \centering
    \includegraphics[width=0.95\linewidth]{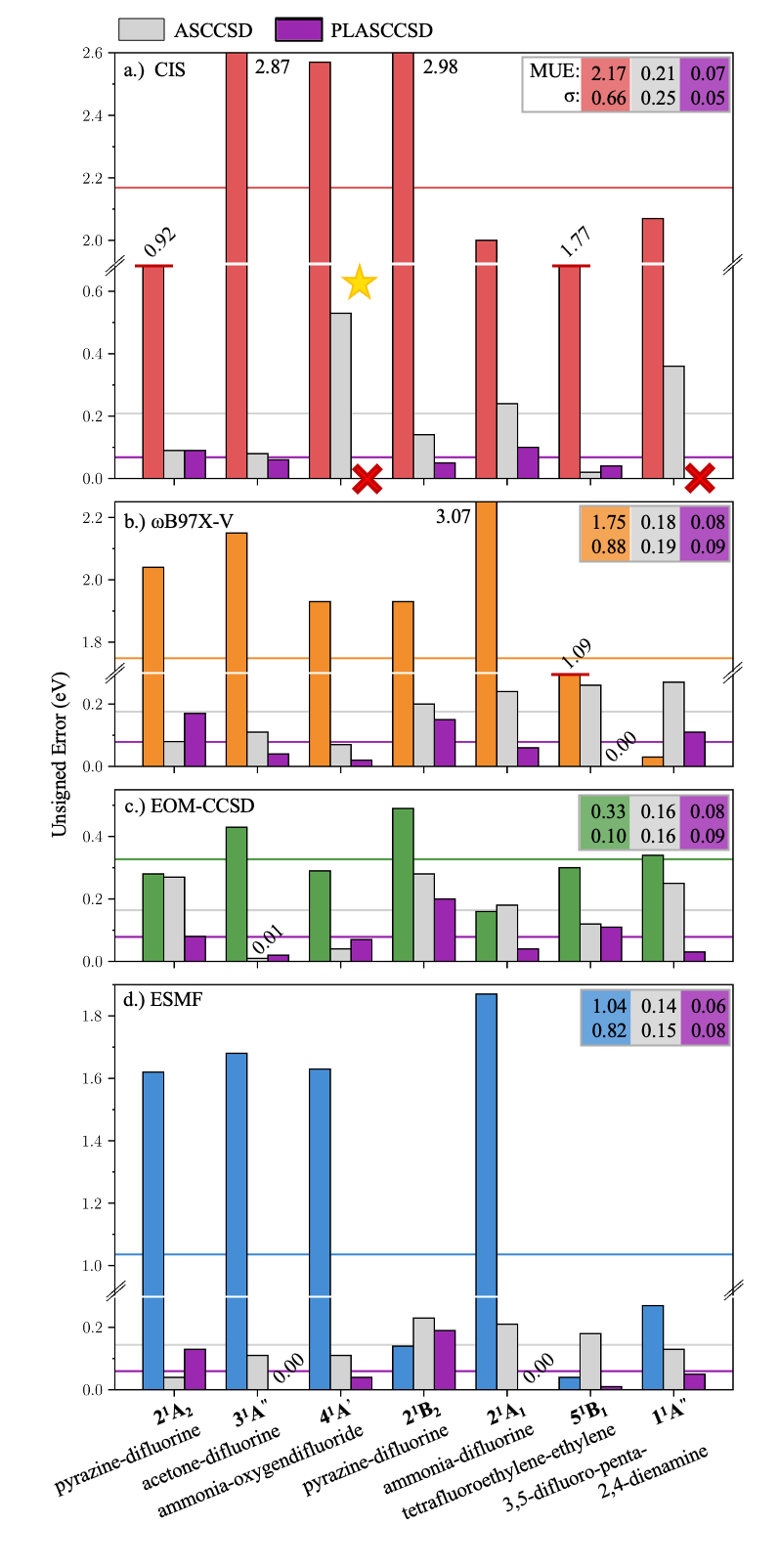}
    \caption{Unsigned errors in charge transfer excitation energies.
             Errors for the starting point methods are shown in
             red (CIS), orange (TD-DFT/$\omega$B97X-V),
             green (EOM-CCSD), and blue (ESMF).
             Errors for ASCC and PLASCC based on these starting
             points are shown in gray and purple, respectively.
             Summarizing statistics across these seven states
             are provided in the insets, and horizontal lines
             depict the MUEs.
             Note the breaks in the vertical axes, and that
             we print the number for the unsigned error at
             the top of bars that do not fit on the plot.
             Red ``x'' marks indicate cases in which ASCC
             or PLASCC did not converge, and the yellow star
             indicates the one starting point that was two-CSF.
    }
    \label{fig:CT_barchart}
\end{figure}

In Figure \ref{fig:CT_barchart}, we see that even in cases where
the starting point method makes multi-eV errors in predicting
a charge transfer energy, ASCC and especially PLASCC
calculations based on that starting point tend to be quite
accurate.
Although the CIS starting point proved to be poor enough to
prevent PLASCC from converging in two cases,
post-linear-response PLASCC makes dramatic improvements
in the other five CIS cases and in  six of the seven TD-DFT cases.
Overall, when PLASCC is started from TD-DFT, it produces a
MUE that is 20 times smaller than that of TD-DFT itself,
which is especially remarkable given that the functional
in question is a range-separated hybrid with dispersion
corrections and thus is relatively well prepared for
charge transfers between the molecules of a dimer.
The improvement when starting from EOM-CCSD is less
dramatic, but still significant, with the MUE lowered
by a factor of more than three.
As in the valence and Rydberg states, starting from ESMF
provides a small advantage, but it is far from clear that
reducing the MUE by 0.02 eV is worth forgoing the
convenience of setting up the starting point via
linear response theory.

\section{Conclusion}
\label{sec:Conclusion}

We have found that Aufbau suppressed coupled cluster is quite
insensitive to its starting point, which allows it to produce
high-accuracy predictions when used in a post-linear-response
mode with a range of linear response starting points.
Even in cases where the underlying linear response theory
is inaccurate --- such as when modeling charge transfer states
with time-dependent density functional theory ---
the post-linear-response coupled cluster treatment
tends to produce accurate results.
While starting from excited state mean field does appear
to offer a (very) small accuracy boost,
the fact that that method must perform a nonlinear orbital
optimization one state at a time suggests that, for many practical
applications, the more easily generated linear response
starting points may be preferred.

Looking forward, there appear to be multiple ways to exploit
the approach's insensitivity to its starting point.
Although we have not tested it here, the CC2 method
now seems quite promising as a starting point,
given that its $N^5$ cost scaling would go well with
our recently-developed strategy of nesting a small
excited state coupled cluster evaluation within
its own non-iterative $N^5$ second order perturbation theory.
That said, CC2's cost is iterative $N^5$,
and so one may also wonder whether the difficulties seen
here for the even more affordable configuration interaction
singles method could be addressed via modest,
selective configuration interaction
style expansions of its determinant basis.
For that matter, transition density matrices and natural
orbitals from selective configuration interaction itself
may also be effective.
There is likely also scope to improve the convergence
chances of ASCC and PLASCC for all starting points by
improving their amplitude equation Jacobians.
We look forward to continuing to explore the opportunities
created by Aufbau suppressed coupled cluster's starting
point insensitivity in the future.

\begin{acknowledgments}
This work was supported by the National Science Foundation, Award Number 2320936. Computational work was performed with the LBNL Lawrencium cluster and the Savio computational cluster resource provided by the Berkeley Research
Computing program at the University of California, Berkeley. T.K.Q. and H.T. acknowledge that this material is based upon work supported by the National Science Foundation Graduate Research Fellowship Program under Grant No. DGE 2146752. Any opinions, findings, and conclusions or recommendations expressed in this material are those of the authors and do not necessarily reflect the views of the National Science Foundation.

\end{acknowledgments}

\section*{Data Availability Statement}

The data that support the findings of this study are available
within the article and its supplementary material.

\section*{References}
\bibliographystyle{achemso}
\bibliography{main}

\providecommand{\latin}[1]{#1}
\makeatletter
\providecommand{\doi}
  {\begingroup\let\do\@makeother\dospecials
  \catcode`\{=1 \catcode`\}=2 \doi@aux}
\providecommand{\doi@aux}[1]{\endgroup\texttt{#1}}
\makeatother
\providecommand*\mcitethebibliography{\thebibliography}
\csname @ifundefined\endcsname{endmcitethebibliography}  {\let\endmcitethebibliography\endthebibliography}{}
\begin{mcitethebibliography}{98}
\providecommand*\natexlab[1]{#1}
\providecommand*\mciteSetBstSublistMode[1]{}
\providecommand*\mciteSetBstMaxWidthForm[2]{}
\providecommand*\mciteBstWouldAddEndPuncttrue
  {\def\EndOfBibitem{\unskip.}}
\providecommand*\mciteBstWouldAddEndPunctfalse
  {\let\EndOfBibitem\relax}
\providecommand*\mciteSetBstMidEndSepPunct[3]{}
\providecommand*\mciteSetBstSublistLabelBeginEnd[3]{}
\providecommand*\EndOfBibitem{}
\mciteSetBstSublistMode{f}
\mciteSetBstMaxWidthForm{subitem}{(\alph{mcitesubitemcount})}
\mciteSetBstSublistLabelBeginEnd
  {\mcitemaxwidthsubitemform\space}
  {\relax}
  {\relax}

\bibitem[Shea and Neuscamman(2018)Shea, and Neuscamman]{shea_communication_2018}
Shea,~J. A.~R.; Neuscamman,~E. Communication: {A} mean field platform for excited state quantum chemistry. \emph{The Journal of Chemical Physics} \textbf{2018}, \emph{149}, 081101\relax
\mciteBstWouldAddEndPuncttrue
\mciteSetBstMidEndSepPunct{\mcitedefaultmidpunct}
{\mcitedefaultendpunct}{\mcitedefaultseppunct}\relax
\EndOfBibitem
\bibitem[Shea \latin{et~al.}(2020)Shea, Gwin, and Neuscamman]{shea_generalized_2020}
Shea,~J. A.~R.; Gwin,~E.; Neuscamman,~E. A {Generalized} {Variational} {Principle} with {Applications} to {Excited} {State} {Mean} {Field} {Theory}. \emph{Journal of Chemical Theory and Computation} \textbf{2020}, \emph{16}, 1526--1540, Publisher: American Chemical Society\relax
\mciteBstWouldAddEndPuncttrue
\mciteSetBstMidEndSepPunct{\mcitedefaultmidpunct}
{\mcitedefaultendpunct}{\mcitedefaultseppunct}\relax
\EndOfBibitem
\bibitem[Hardikar and Neuscamman(2020)Hardikar, and Neuscamman]{hardikar_self-consistent_2020}
Hardikar,~T.~S.; Neuscamman,~E. A self-consistent field formulation of excited state mean field theory. \emph{The Journal of Chemical Physics} \textbf{2020}, \emph{153}, 164108\relax
\mciteBstWouldAddEndPuncttrue
\mciteSetBstMidEndSepPunct{\mcitedefaultmidpunct}
{\mcitedefaultendpunct}{\mcitedefaultseppunct}\relax
\EndOfBibitem
\bibitem[Hait and Head-Gordon(2020)Hait, and Head-Gordon]{hait2020excited}
Hait,~D.; Head-Gordon,~M. Excited state orbital optimization via minimizing the square of the gradient: General approach and application to singly and doubly excited states via density functional theory. \emph{Journal of chemical theory and computation} \textbf{2020}, \emph{16}, 1699--1710\relax
\mciteBstWouldAddEndPuncttrue
\mciteSetBstMidEndSepPunct{\mcitedefaultmidpunct}
{\mcitedefaultendpunct}{\mcitedefaultseppunct}\relax
\EndOfBibitem
\bibitem[Hait and Head-Gordon(2020)Hait, and Head-Gordon]{hait2020highly}
Hait,~D.; Head-Gordon,~M. Highly accurate prediction of core spectra of molecules at density functional theory cost: Attaining sub-electronvolt error from a restricted open-shell Kohn--Sham approach. \emph{The journal of physical chemistry letters} \textbf{2020}, \emph{11}, 775--786\relax
\mciteBstWouldAddEndPuncttrue
\mciteSetBstMidEndSepPunct{\mcitedefaultmidpunct}
{\mcitedefaultendpunct}{\mcitedefaultseppunct}\relax
\EndOfBibitem
\bibitem[Hait and Head-Gordon(2021)Hait, and Head-Gordon]{hait2021orbital}
Hait,~D.; Head-Gordon,~M. Orbital optimized density functional theory for electronic excited states. \emph{The journal of physical chemistry letters} \textbf{2021}, \emph{12}, 4517--4529\relax
\mciteBstWouldAddEndPuncttrue
\mciteSetBstMidEndSepPunct{\mcitedefaultmidpunct}
{\mcitedefaultendpunct}{\mcitedefaultseppunct}\relax
\EndOfBibitem
\bibitem[Tran \latin{et~al.}(2019)Tran, Shea, and Neuscamman]{tran2019tracking}
Tran,~L.~N.; Shea,~J.~A.; Neuscamman,~E. Tracking excited states in wave function optimization using density matrices and variational principles. \emph{Journal of chemical theory and computation} \textbf{2019}, \emph{15}, 4790--4803\relax
\mciteBstWouldAddEndPuncttrue
\mciteSetBstMidEndSepPunct{\mcitedefaultmidpunct}
{\mcitedefaultendpunct}{\mcitedefaultseppunct}\relax
\EndOfBibitem
\bibitem[Tran and Neuscamman(2020)Tran, and Neuscamman]{tran2020improving}
Tran,~L.~N.; Neuscamman,~E. Improving excited-state potential energy surfaces via optimal orbital shapes. \emph{The Journal of Physical Chemistry A} \textbf{2020}, \emph{124}, 8273--8279\relax
\mciteBstWouldAddEndPuncttrue
\mciteSetBstMidEndSepPunct{\mcitedefaultmidpunct}
{\mcitedefaultendpunct}{\mcitedefaultseppunct}\relax
\EndOfBibitem
\bibitem[Tran and Neuscamman(2023)Tran, and Neuscamman]{tran2023exploring}
Tran,~L.~N.; Neuscamman,~E. Exploring Ligand-to-Metal Charge-Transfer States in the Photo-Ferrioxalate System Using Excited-State Specific Optimization. \emph{The Journal of Physical Chemistry Letters} \textbf{2023}, \emph{14}, 7454--7460\relax
\mciteBstWouldAddEndPuncttrue
\mciteSetBstMidEndSepPunct{\mcitedefaultmidpunct}
{\mcitedefaultendpunct}{\mcitedefaultseppunct}\relax
\EndOfBibitem
\bibitem[Garner \latin{et~al.}(2024)Garner, Haugen, Leone, and Neuscamman]{garner2024spin}
Garner,~S.~M.; Haugen,~E.~A.; Leone,~S.~R.; Neuscamman,~E. Spin coupling effect on geometry-dependent x-ray absorption of diradicals. \emph{Journal of the American Chemical Society} \textbf{2024}, \emph{146}, 2387--2397\relax
\mciteBstWouldAddEndPuncttrue
\mciteSetBstMidEndSepPunct{\mcitedefaultmidpunct}
{\mcitedefaultendpunct}{\mcitedefaultseppunct}\relax
\EndOfBibitem
\bibitem[Crawford and Schaefer~III(2007)Crawford, and Schaefer~III]{crawford2007introduction}
Crawford,~T.~D.; Schaefer~III,~H.~F. An introduction to coupled cluster theory for computational chemists. \emph{Reviews in computational chemistry} \textbf{2007}, \emph{14}, 33--136\relax
\mciteBstWouldAddEndPuncttrue
\mciteSetBstMidEndSepPunct{\mcitedefaultmidpunct}
{\mcitedefaultendpunct}{\mcitedefaultseppunct}\relax
\EndOfBibitem
\bibitem[Bartlett and Musia{\l}(2007)Bartlett, and Musia{\l}]{bartlett2007coupled}
Bartlett,~R.~J.; Musia{\l},~M. Coupled-cluster theory in quantum chemistry. \emph{Reviews of Modern Physics} \textbf{2007}, \emph{79}, 291\relax
\mciteBstWouldAddEndPuncttrue
\mciteSetBstMidEndSepPunct{\mcitedefaultmidpunct}
{\mcitedefaultendpunct}{\mcitedefaultseppunct}\relax
\EndOfBibitem
\bibitem[Shavitt and Bartlett(2009)Shavitt, and Bartlett]{shavitt2009many}
Shavitt,~I.; Bartlett,~R.~J. \emph{Many-body methods in chemistry and physics: MBPT and coupled-cluster theory}; Cambridge university press, 2009\relax
\mciteBstWouldAddEndPuncttrue
\mciteSetBstMidEndSepPunct{\mcitedefaultmidpunct}
{\mcitedefaultendpunct}{\mcitedefaultseppunct}\relax
\EndOfBibitem
\bibitem[Hampel \latin{et~al.}(1992)Hampel, Peterson, and Werner]{hampel1992comparison}
Hampel,~C.; Peterson,~K.~A.; Werner,~H.-J. A comparison of the efficiency and accuracy of the quadratic configuration interaction (QCISD), coupled cluster (CCSD), and Brueckner coupled cluster (BCCD) methods. \emph{Chemical physics letters} \textbf{1992}, \emph{190}, 1--12\relax
\mciteBstWouldAddEndPuncttrue
\mciteSetBstMidEndSepPunct{\mcitedefaultmidpunct}
{\mcitedefaultendpunct}{\mcitedefaultseppunct}\relax
\EndOfBibitem
\bibitem[Lee \latin{et~al.}(1992)Lee, Kobayashi, Handy, and Amos]{lee1992comparison}
Lee,~T.~J.; Kobayashi,~R.; Handy,~N.~C.; Amos,~R.~D. Comparison of the Brueckner and coupled-cluster approaches to electron correlation. \emph{The Journal of chemical physics} \textbf{1992}, \emph{96}, 8931--8937\relax
\mciteBstWouldAddEndPuncttrue
\mciteSetBstMidEndSepPunct{\mcitedefaultmidpunct}
{\mcitedefaultendpunct}{\mcitedefaultseppunct}\relax
\EndOfBibitem
\bibitem[Bertels \latin{et~al.}(2021)Bertels, Lee, and Head-Gordon]{bertels2021polishing}
Bertels,~L.~W.; Lee,~J.; Head-Gordon,~M. Polishing the gold standard: The role of orbital choice in CCSD (T) vibrational frequency prediction. \emph{Journal of chemical theory and computation} \textbf{2021}, \emph{17}, 742--755\relax
\mciteBstWouldAddEndPuncttrue
\mciteSetBstMidEndSepPunct{\mcitedefaultmidpunct}
{\mcitedefaultendpunct}{\mcitedefaultseppunct}\relax
\EndOfBibitem
\bibitem[Tuckman and Neuscamman(2024)Tuckman, and Neuscamman]{tuckman_aufbau_2024}
Tuckman,~H.; Neuscamman,~E. Aufbau {Suppressed} {Coupled} {Cluster} {Theory} for {Electronically} {Excited} {States}. \emph{Journal of Chemical Theory and Computation} \textbf{2024}, \emph{20}, 2761--2773, Publisher: American Chemical Society\relax
\mciteBstWouldAddEndPuncttrue
\mciteSetBstMidEndSepPunct{\mcitedefaultmidpunct}
{\mcitedefaultendpunct}{\mcitedefaultseppunct}\relax
\EndOfBibitem
\bibitem[Tuckman \latin{et~al.}(2025)Tuckman, Ma, and Neuscamman]{tuckman_improving_2025}
Tuckman,~H.; Ma,~Z.; Neuscamman,~E. Improving {Aufbau} {Suppressed} {Coupled} {Cluster} through {Perturbative} {Analysis}. \emph{Journal of Chemical Theory and Computation} \textbf{2025}, Publisher: American Chemical Society\relax
\mciteBstWouldAddEndPuncttrue
\mciteSetBstMidEndSepPunct{\mcitedefaultmidpunct}
{\mcitedefaultendpunct}{\mcitedefaultseppunct}\relax
\EndOfBibitem
\bibitem[Tuckman and Neuscamman(2025)Tuckman, and Neuscamman]{tuckman2025NestingArXiv}
Tuckman,~H.; Neuscamman,~E. Fast and Accurate Charge Transfer Excitations via Nested Aufbau Suppressed Coupled Cluster. \emph{arXiv.org} \textbf{2025}, 2505.17299\relax
\mciteBstWouldAddEndPuncttrue
\mciteSetBstMidEndSepPunct{\mcitedefaultmidpunct}
{\mcitedefaultendpunct}{\mcitedefaultseppunct}\relax
\EndOfBibitem
\bibitem[Thouless(1960)]{thouless1960stability}
Thouless,~D.~J. Stability conditions and nuclear rotations in the Hartree-Fock theory. \emph{Nuclear Physics} \textbf{1960}, \emph{21}, 225--232\relax
\mciteBstWouldAddEndPuncttrue
\mciteSetBstMidEndSepPunct{\mcitedefaultmidpunct}
{\mcitedefaultendpunct}{\mcitedefaultseppunct}\relax
\EndOfBibitem
\bibitem[Dreuw and Head-Gordon(2005)Dreuw, and Head-Gordon]{dreuw2005single}
Dreuw,~A.; Head-Gordon,~M. Single-reference ab initio methods for the calculation of excited states of large molecules. \emph{Chemical reviews} \textbf{2005}, \emph{105}, 4009--4037\relax
\mciteBstWouldAddEndPuncttrue
\mciteSetBstMidEndSepPunct{\mcitedefaultmidpunct}
{\mcitedefaultendpunct}{\mcitedefaultseppunct}\relax
\EndOfBibitem
\bibitem[Runge and Gross(1984)Runge, and Gross]{runge1984density}
Runge,~E.; Gross,~E.~K. Density-functional theory for time-dependent systems. \emph{Physical review letters} \textbf{1984}, \emph{52}, 997\relax
\mciteBstWouldAddEndPuncttrue
\mciteSetBstMidEndSepPunct{\mcitedefaultmidpunct}
{\mcitedefaultendpunct}{\mcitedefaultseppunct}\relax
\EndOfBibitem
\bibitem[Burke \latin{et~al.}(2005)Burke, Werschnik, and Gross]{burke2005time}
Burke,~K.; Werschnik,~J.; Gross,~E. Time-dependent density functional theory: Past, present, and future. \emph{The Journal of chemical physics} \textbf{2005}, \emph{123}, 062206\relax
\mciteBstWouldAddEndPuncttrue
\mciteSetBstMidEndSepPunct{\mcitedefaultmidpunct}
{\mcitedefaultendpunct}{\mcitedefaultseppunct}\relax
\EndOfBibitem
\bibitem[Casida and Huix-Rotllant(2012)Casida, and Huix-Rotllant]{casida2012progress}
Casida,~M.~E.; Huix-Rotllant,~M. Progress in time-dependent density-functional theory. \emph{Annual review of physical chemistry} \textbf{2012}, \emph{63}, 287--323\relax
\mciteBstWouldAddEndPuncttrue
\mciteSetBstMidEndSepPunct{\mcitedefaultmidpunct}
{\mcitedefaultendpunct}{\mcitedefaultseppunct}\relax
\EndOfBibitem
\bibitem[Rowe(1968)]{rowe1968equations}
Rowe,~D. Equations-of-motion method and the extended shell model. \emph{Reviews of Modern Physics} \textbf{1968}, \emph{40}, 153\relax
\mciteBstWouldAddEndPuncttrue
\mciteSetBstMidEndSepPunct{\mcitedefaultmidpunct}
{\mcitedefaultendpunct}{\mcitedefaultseppunct}\relax
\EndOfBibitem
\bibitem[Stanton and Bartlett(1993)Stanton, and Bartlett]{stanton1993equation}
Stanton,~J.~F.; Bartlett,~R.~J. The equation of motion coupled-cluster method. A systematic biorthogonal approach to molecular excitation energies, transition probabilities, and excited state properties. \emph{The Journal of chemical physics} \textbf{1993}, \emph{98}, 7029--7039\relax
\mciteBstWouldAddEndPuncttrue
\mciteSetBstMidEndSepPunct{\mcitedefaultmidpunct}
{\mcitedefaultendpunct}{\mcitedefaultseppunct}\relax
\EndOfBibitem
\bibitem[Krylov(2008)]{krylov2008equation}
Krylov,~A.~I. Equation-of-motion coupled-cluster methods for open-shell and electronically excited species: The hitchhiker's guide to Fock space. \emph{Annu. Rev. Phys. Chem.} \textbf{2008}, \emph{59}, 433--462\relax
\mciteBstWouldAddEndPuncttrue
\mciteSetBstMidEndSepPunct{\mcitedefaultmidpunct}
{\mcitedefaultendpunct}{\mcitedefaultseppunct}\relax
\EndOfBibitem
\bibitem[Monkhorst(1977)]{monkhorst1977calculation}
Monkhorst,~H.~J. Calculation of properties with the coupled-cluster method. \emph{International Journal of Quantum Chemistry} \textbf{1977}, \emph{12}, 421--432\relax
\mciteBstWouldAddEndPuncttrue
\mciteSetBstMidEndSepPunct{\mcitedefaultmidpunct}
{\mcitedefaultendpunct}{\mcitedefaultseppunct}\relax
\EndOfBibitem
\bibitem[Dalgaard and Monkhorst(1983)Dalgaard, and Monkhorst]{dalgaard1983some}
Dalgaard,~E.; Monkhorst,~H.~J. Some aspects of the time-dependent coupled-cluster approach to dynamic response functions. \emph{Physical Review A} \textbf{1983}, \emph{28}, 1217\relax
\mciteBstWouldAddEndPuncttrue
\mciteSetBstMidEndSepPunct{\mcitedefaultmidpunct}
{\mcitedefaultendpunct}{\mcitedefaultseppunct}\relax
\EndOfBibitem
\bibitem[Sekino and Bartlett(1984)Sekino, and Bartlett]{sekino1984linear}
Sekino,~H.; Bartlett,~R.~J. A linear response, coupled-cluster theory for excitation energy. \emph{International Journal of Quantum Chemistry} \textbf{1984}, \emph{26}, 255--265\relax
\mciteBstWouldAddEndPuncttrue
\mciteSetBstMidEndSepPunct{\mcitedefaultmidpunct}
{\mcitedefaultendpunct}{\mcitedefaultseppunct}\relax
\EndOfBibitem
\bibitem[Koch and J{\o}rgensen(1990)Koch, and J{\o}rgensen]{koch1990coupled}
Koch,~H.; J{\o}rgensen,~P. Coupled cluster response functions. \emph{The Journal of chemical physics} \textbf{1990}, \emph{93}, 3333\relax
\mciteBstWouldAddEndPuncttrue
\mciteSetBstMidEndSepPunct{\mcitedefaultmidpunct}
{\mcitedefaultendpunct}{\mcitedefaultseppunct}\relax
\EndOfBibitem
\bibitem[Rico and Head-Gordon(1993)Rico, and Head-Gordon]{rico1993single}
Rico,~R.~J.; Head-Gordon,~M. Single-reference theories of molecular excited states with single and double substitutions. \emph{Chemical physics letters} \textbf{1993}, \emph{213}, 224--232\relax
\mciteBstWouldAddEndPuncttrue
\mciteSetBstMidEndSepPunct{\mcitedefaultmidpunct}
{\mcitedefaultendpunct}{\mcitedefaultseppunct}\relax
\EndOfBibitem
\bibitem[Koch \latin{et~al.}(1994)Koch, Kobayashi, Sanchez~de Mer{\'a}s, and J{\o}rgensen]{koch1994calculation}
Koch,~H.; Kobayashi,~R.; Sanchez~de Mer{\'a}s,~A.; J{\o}rgensen,~P. Calculation of size-intensive transition moments from the coupled cluster singles and doubles linear response function. \emph{The Journal of chemical physics} \textbf{1994}, \emph{100}, 4393--4400\relax
\mciteBstWouldAddEndPuncttrue
\mciteSetBstMidEndSepPunct{\mcitedefaultmidpunct}
{\mcitedefaultendpunct}{\mcitedefaultseppunct}\relax
\EndOfBibitem
\bibitem[Sneskov and Christiansen(2012)Sneskov, and Christiansen]{sneskov2012excited}
Sneskov,~K.; Christiansen,~O. Excited state coupled cluster methods. \emph{Wiley Interdisciplinary Reviews: Computational Molecular Science} \textbf{2012}, \emph{2}, 566--584\relax
\mciteBstWouldAddEndPuncttrue
\mciteSetBstMidEndSepPunct{\mcitedefaultmidpunct}
{\mcitedefaultendpunct}{\mcitedefaultseppunct}\relax
\EndOfBibitem
\bibitem[Subotnik(2011)]{subotnik2011communication}
Subotnik,~J.~E. Communication: Configuration interaction singles has a large systematic bias against charge-transfer states. \emph{The Journal of chemical physics} \textbf{2011}, \emph{135}, 071104\relax
\mciteBstWouldAddEndPuncttrue
\mciteSetBstMidEndSepPunct{\mcitedefaultmidpunct}
{\mcitedefaultendpunct}{\mcitedefaultseppunct}\relax
\EndOfBibitem
\bibitem[Herbert(2023)]{herbert2023density}
Herbert,~J.~M. \emph{Theoretical and computational photochemistry}; Elsevier, 2023; pp 69--118\relax
\mciteBstWouldAddEndPuncttrue
\mciteSetBstMidEndSepPunct{\mcitedefaultmidpunct}
{\mcitedefaultendpunct}{\mcitedefaultseppunct}\relax
\EndOfBibitem
\bibitem[Sobolewski and Domcke(2003)Sobolewski, and Domcke]{sobolewski2003ab}
Sobolewski,~A.~L.; Domcke,~W. Ab initio study of the excited-state coupled electron--proton-transfer process in the 2-aminopyridine dimer. \emph{Chemical Physics} \textbf{2003}, \emph{294}, 73--83\relax
\mciteBstWouldAddEndPuncttrue
\mciteSetBstMidEndSepPunct{\mcitedefaultmidpunct}
{\mcitedefaultendpunct}{\mcitedefaultseppunct}\relax
\EndOfBibitem
\bibitem[Dreuw \latin{et~al.}(2003)Dreuw, Weisman, and Head-Gordon]{dreuw2003long}
Dreuw,~A.; Weisman,~J.~L.; Head-Gordon,~M. Long-range charge-transfer excited states in time-dependent density functional theory require non-local exchange. \emph{The Journal of chemical physics} \textbf{2003}, \emph{119}, 2943--2946\relax
\mciteBstWouldAddEndPuncttrue
\mciteSetBstMidEndSepPunct{\mcitedefaultmidpunct}
{\mcitedefaultendpunct}{\mcitedefaultseppunct}\relax
\EndOfBibitem
\bibitem[Dreuw and Head-Gordon(2004)Dreuw, and Head-Gordon]{dreuw2004failure}
Dreuw,~A.; Head-Gordon,~M. Failure of time-dependent density functional theory for long-range charge-transfer excited states: the zincbacteriochlorin- bacteriochlorin and bacteriochlorophyll- spheroidene complexes. \emph{Journal of the American Chemical Society} \textbf{2004}, \emph{126}, 4007--4016\relax
\mciteBstWouldAddEndPuncttrue
\mciteSetBstMidEndSepPunct{\mcitedefaultmidpunct}
{\mcitedefaultendpunct}{\mcitedefaultseppunct}\relax
\EndOfBibitem
\bibitem[Mester and K{\'a}llay(2022)Mester, and K{\'a}llay]{mester2022charge}
Mester,~D.; K{\'a}llay,~M. Charge-transfer excitations within density functional theory: how accurate are the most recommended approaches? \emph{Journal of Chemical Theory and Computation} \textbf{2022}, \emph{18}, 1646--1662\relax
\mciteBstWouldAddEndPuncttrue
\mciteSetBstMidEndSepPunct{\mcitedefaultmidpunct}
{\mcitedefaultendpunct}{\mcitedefaultseppunct}\relax
\EndOfBibitem
\bibitem[Kozma \latin{et~al.}(2020)Kozma, Tajti, Demoulin, Izs{\'a}k, Nooijen, and Szalay]{kozma2020new}
Kozma,~B.; Tajti,~A.; Demoulin,~B.; Izs{\'a}k,~R.; Nooijen,~M.; Szalay,~P.~G. A new benchmark set for excitation energy of charge transfer states: systematic investigation of coupled cluster type methods. \emph{Journal of Chemical Theory and Computation} \textbf{2020}, \emph{16}, 4213--4225\relax
\mciteBstWouldAddEndPuncttrue
\mciteSetBstMidEndSepPunct{\mcitedefaultmidpunct}
{\mcitedefaultendpunct}{\mcitedefaultseppunct}\relax
\EndOfBibitem
\bibitem[Izs{\'a}k(2020)]{izsak2020single}
Izs{\'a}k,~R. Single-reference coupled cluster methods for computing excitation energies in large molecules: The efficiency and accuracy of approximations. \emph{Wiley Interdisciplinary Reviews: Computational Molecular Science} \textbf{2020}, \emph{10}, e1445\relax
\mciteBstWouldAddEndPuncttrue
\mciteSetBstMidEndSepPunct{\mcitedefaultmidpunct}
{\mcitedefaultendpunct}{\mcitedefaultseppunct}\relax
\EndOfBibitem
\bibitem[Coriani \latin{et~al.}(2012)Coriani, Christiansen, Fransson, and Norman]{coriani2012coupled}
Coriani,~S.; Christiansen,~O.; Fransson,~T.; Norman,~P. Coupled-cluster response theory for near-edge x-ray-absorption fine structure of atoms and molecules. \emph{Physical Review A—Atomic, Molecular, and Optical Physics} \textbf{2012}, \emph{85}, 022507\relax
\mciteBstWouldAddEndPuncttrue
\mciteSetBstMidEndSepPunct{\mcitedefaultmidpunct}
{\mcitedefaultendpunct}{\mcitedefaultseppunct}\relax
\EndOfBibitem
\bibitem[Frati \latin{et~al.}(2019)Frati, De~Groot, Cerezo, Santoro, Cheng, Faber, and Coriani]{frati2019coupled}
Frati,~F.; De~Groot,~F.; Cerezo,~J.; Santoro,~F.; Cheng,~L.; Faber,~R.; Coriani,~S. Coupled cluster study of the x-ray absorption spectra of formaldehyde derivatives at the oxygen, carbon, and fluorine K-edges. \emph{The Journal of Chemical Physics} \textbf{2019}, \emph{151}, 064107\relax
\mciteBstWouldAddEndPuncttrue
\mciteSetBstMidEndSepPunct{\mcitedefaultmidpunct}
{\mcitedefaultendpunct}{\mcitedefaultseppunct}\relax
\EndOfBibitem
\bibitem[Domingo \latin{et~al.}(2012)Domingo, Carvajal, de~Graaf, Sivalingam, Neese, and Angeli]{domingo2012metal}
Domingo,~A.; Carvajal,~M.~{\`A}.; de~Graaf,~C.; Sivalingam,~K.; Neese,~F.; Angeli,~C. Metal-to-metal charge-transfer transitions: reliable excitation energies from ab initio calculations. \emph{Theor. Chem. Acc.} \textbf{2012}, \emph{131}, 1264\relax
\mciteBstWouldAddEndPuncttrue
\mciteSetBstMidEndSepPunct{\mcitedefaultmidpunct}
{\mcitedefaultendpunct}{\mcitedefaultseppunct}\relax
\EndOfBibitem
\bibitem[Pineda~Flores and Neuscamman(2019)Pineda~Flores, and Neuscamman]{pineda2019excited}
Pineda~Flores,~S.~D.; Neuscamman,~E. Excited state specific multi-Slater Jastrow wave functions. \emph{The Journal of Physical Chemistry A} \textbf{2019}, \emph{123}, 1487--1497\relax
\mciteBstWouldAddEndPuncttrue
\mciteSetBstMidEndSepPunct{\mcitedefaultmidpunct}
{\mcitedefaultendpunct}{\mcitedefaultseppunct}\relax
\EndOfBibitem
\bibitem[Clune and Neuscamman(2025)Clune, and Neuscamman]{clune2025emlc}
Clune,~R.; Neuscamman,~E. An excitation matched local correlation approach to excited state specific perturbation theory. \emph{arXiv} \textbf{2025}, 2505.08659\relax
\mciteBstWouldAddEndPuncttrue
\mciteSetBstMidEndSepPunct{\mcitedefaultmidpunct}
{\mcitedefaultendpunct}{\mcitedefaultseppunct}\relax
\EndOfBibitem
\bibitem[Ye \latin{et~al.}(2017)Ye, Welborn, Ricke, and Van~Voorhis]{ye2017sigma}
Ye,~H.-Z.; Welborn,~M.; Ricke,~N.~D.; Van~Voorhis,~T. $\sigma$-SCF: A direct energy-targeting method to mean-field excited states. \emph{The Journal of chemical physics} \textbf{2017}, \emph{147}, 214104\relax
\mciteBstWouldAddEndPuncttrue
\mciteSetBstMidEndSepPunct{\mcitedefaultmidpunct}
{\mcitedefaultendpunct}{\mcitedefaultseppunct}\relax
\EndOfBibitem
\bibitem[Ye and Van~Voorhis(2019)Ye, and Van~Voorhis]{ye2019half}
Ye,~H.-Z.; Van~Voorhis,~T. Half-projected $\sigma$ self-consistent field for electronic excited states. \emph{Journal of chemical theory and computation} \textbf{2019}, \emph{15}, 2954--2965\relax
\mciteBstWouldAddEndPuncttrue
\mciteSetBstMidEndSepPunct{\mcitedefaultmidpunct}
{\mcitedefaultendpunct}{\mcitedefaultseppunct}\relax
\EndOfBibitem
\bibitem[Bagus(1965)]{bagus1965scf}
Bagus,~P.~S. Self-consistent-field wave functions for hole states of some Ne-like and Ar-like ions. \emph{Phys. Rev.} \textbf{1965}, \emph{139}, A619\relax
\mciteBstWouldAddEndPuncttrue
\mciteSetBstMidEndSepPunct{\mcitedefaultmidpunct}
{\mcitedefaultendpunct}{\mcitedefaultseppunct}\relax
\EndOfBibitem
\bibitem[Hsu \latin{et~al.}(1976)Hsu, Davidson, and Pitzer]{Pitzer1976scf}
Hsu,~H.-l.; Davidson,~E.~R.; Pitzer,~R.~M. An SCF method for hole states. \emph{J. Chem. Phys.} \textbf{1976}, \emph{65}, 609--613\relax
\mciteBstWouldAddEndPuncttrue
\mciteSetBstMidEndSepPunct{\mcitedefaultmidpunct}
{\mcitedefaultendpunct}{\mcitedefaultseppunct}\relax
\EndOfBibitem
\bibitem[Naves~de Brito \latin{et~al.}(1991)Naves~de Brito, Correia, Svensson, and {\AA}gren]{argen1991xray}
Naves~de Brito,~A.; Correia,~N.; Svensson,~S.; {\AA}gren,~H. A theoretical study of x-ray photoelectron spectra of model molecules for polymethylmethacrylate. \emph{J. Chem. Phys.} \textbf{1991}, \emph{95}, 2965--2974\relax
\mciteBstWouldAddEndPuncttrue
\mciteSetBstMidEndSepPunct{\mcitedefaultmidpunct}
{\mcitedefaultendpunct}{\mcitedefaultseppunct}\relax
\EndOfBibitem
\bibitem[Besley \latin{et~al.}(2009)Besley, Gilbert, and Gill]{besley2009self}
Besley,~N.~A.; Gilbert,~A.~T.; Gill,~P.~M. Self-consistent-field calculations of core excited states. \emph{The Journal of chemical physics} \textbf{2009}, \emph{130}, 124308\relax
\mciteBstWouldAddEndPuncttrue
\mciteSetBstMidEndSepPunct{\mcitedefaultmidpunct}
{\mcitedefaultendpunct}{\mcitedefaultseppunct}\relax
\EndOfBibitem
\bibitem[Filatov and Shaik(1999)Filatov, and Shaik]{Shaik1999}
Filatov,~M.; Shaik,~S. A spin-restricted ensemble-referenced Kohn-Sham method and its application to diradicaloid situations. \emph{Chem. Phys. Lett.} \textbf{1999}, \emph{304}, 429--437\relax
\mciteBstWouldAddEndPuncttrue
\mciteSetBstMidEndSepPunct{\mcitedefaultmidpunct}
{\mcitedefaultendpunct}{\mcitedefaultseppunct}\relax
\EndOfBibitem
\bibitem[Kowalczyk \latin{et~al.}(2013)Kowalczyk, Tsuchimochi, Chen, Top, and Van~Voorhis]{kowalczyk2013excitation}
Kowalczyk,~T.; Tsuchimochi,~T.; Chen,~P.-T.; Top,~L.; Van~Voorhis,~T. Excitation energies and Stokes shifts from a restricted open-shell Kohn-Sham approach. \emph{The Journal of chemical physics} \textbf{2013}, \emph{138}, 164101\relax
\mciteBstWouldAddEndPuncttrue
\mciteSetBstMidEndSepPunct{\mcitedefaultmidpunct}
{\mcitedefaultendpunct}{\mcitedefaultseppunct}\relax
\EndOfBibitem
\bibitem[Kowalczyk \latin{et~al.}(2011)Kowalczyk, Yost, and Voorhis]{kowalczyk2011assessment}
Kowalczyk,~T.; Yost,~S.~R.; Voorhis,~T.~V. Assessment of the $\Delta$SCF density functional theory approach for electronic excitations in organic dyes. \emph{The Journal of chemical physics} \textbf{2011}, \emph{134}\relax
\mciteBstWouldAddEndPuncttrue
\mciteSetBstMidEndSepPunct{\mcitedefaultmidpunct}
{\mcitedefaultendpunct}{\mcitedefaultseppunct}\relax
\EndOfBibitem
\bibitem[Zhao and Neuscamman(2019)Zhao, and Neuscamman]{zhao2019density}
Zhao,~L.; Neuscamman,~E. Density functional extension to excited-state mean-field theory. \emph{Journal of chemical theory and computation} \textbf{2019}, \emph{16}, 164--178\relax
\mciteBstWouldAddEndPuncttrue
\mciteSetBstMidEndSepPunct{\mcitedefaultmidpunct}
{\mcitedefaultendpunct}{\mcitedefaultseppunct}\relax
\EndOfBibitem
\bibitem[Levi \latin{et~al.}(2020)Levi, Ivanov, and J{\'o}nsson]{levi2020variational}
Levi,~G.; Ivanov,~A.~V.; J{\'o}nsson,~H. Variational density functional calculations of excited states via direct optimization. \emph{Journal of Chemical Theory and Computation} \textbf{2020}, \emph{16}, 6968--6982\relax
\mciteBstWouldAddEndPuncttrue
\mciteSetBstMidEndSepPunct{\mcitedefaultmidpunct}
{\mcitedefaultendpunct}{\mcitedefaultseppunct}\relax
\EndOfBibitem
\bibitem[Kempfer-Robertson \latin{et~al.}(2022)Kempfer-Robertson, Haase, Bersson, Avdic, and Thompson]{kempfer2022role}
Kempfer-Robertson,~E.~M.; Haase,~M.~N.; Bersson,~J.~S.; Avdic,~I.; Thompson,~L.~M. Role of Exact Exchange in Difference Projected Double-Hybrid Density Functional Theory for Treatment of Local, Charge Transfer, and Rydberg Excitations. \emph{The Journal of Physical Chemistry A} \textbf{2022}, \emph{126}, 8058--8069\relax
\mciteBstWouldAddEndPuncttrue
\mciteSetBstMidEndSepPunct{\mcitedefaultmidpunct}
{\mcitedefaultendpunct}{\mcitedefaultseppunct}\relax
\EndOfBibitem
\bibitem[Gilbert \latin{et~al.}(2008)Gilbert, Besley, and Gill]{gilbert2008self}
Gilbert,~A.~T.; Besley,~N.~A.; Gill,~P.~M. Self-consistent field calculations of excited states using the maximum overlap method (MOM). \emph{The Journal of Physical Chemistry A} \textbf{2008}, \emph{112}, 13164--13171\relax
\mciteBstWouldAddEndPuncttrue
\mciteSetBstMidEndSepPunct{\mcitedefaultmidpunct}
{\mcitedefaultendpunct}{\mcitedefaultseppunct}\relax
\EndOfBibitem
\bibitem[Barca \latin{et~al.}(2018)Barca, Gilbert, and Gill]{barca2018simple}
Barca,~G.~M.; Gilbert,~A.~T.; Gill,~P.~M. Simple models for difficult electronic excitations. \emph{Journal of chemical theory and computation} \textbf{2018}, \emph{14}, 1501--1509\relax
\mciteBstWouldAddEndPuncttrue
\mciteSetBstMidEndSepPunct{\mcitedefaultmidpunct}
{\mcitedefaultendpunct}{\mcitedefaultseppunct}\relax
\EndOfBibitem
\bibitem[Carter-Fenk and Herbert(2020)Carter-Fenk, and Herbert]{carter2020state}
Carter-Fenk,~K.; Herbert,~J.~M. State-targeted energy projection: A simple and robust approach to orbital relaxation of non-Aufbau self-consistent field solutions. \emph{Journal of Chemical Theory and Computation} \textbf{2020}, \emph{16}, 5067--5082\relax
\mciteBstWouldAddEndPuncttrue
\mciteSetBstMidEndSepPunct{\mcitedefaultmidpunct}
{\mcitedefaultendpunct}{\mcitedefaultseppunct}\relax
\EndOfBibitem
\bibitem[Zhao and Neuscamman(2016)Zhao, and Neuscamman]{zhao2016efficient}
Zhao,~L.; Neuscamman,~E. An efficient variational principle for the direct optimization of excited states. \emph{Journal of chemical theory and computation} \textbf{2016}, \emph{12}, 3436--3440\relax
\mciteBstWouldAddEndPuncttrue
\mciteSetBstMidEndSepPunct{\mcitedefaultmidpunct}
{\mcitedefaultendpunct}{\mcitedefaultseppunct}\relax
\EndOfBibitem
\bibitem[Robinson \latin{et~al.}(2017)Robinson, Pineda~Flores, and Neuscamman]{robinson2017excitation}
Robinson,~P.~J.; Pineda~Flores,~S.~D.; Neuscamman,~E. Excitation variance matching with limited configuration interaction expansions in variational Monte Carlo. \emph{The Journal of Chemical Physics} \textbf{2017}, \emph{147}, 164114\relax
\mciteBstWouldAddEndPuncttrue
\mciteSetBstMidEndSepPunct{\mcitedefaultmidpunct}
{\mcitedefaultendpunct}{\mcitedefaultseppunct}\relax
\EndOfBibitem
\bibitem[Blunt and Neuscamman(2017)Blunt, and Neuscamman]{blunt2017charge}
Blunt,~N.~S.; Neuscamman,~E. Charge-transfer excited states: Seeking a balanced and efficient wave function ansatz in variational Monte Carlo. \emph{The Journal of Chemical Physics} \textbf{2017}, \emph{147}, 194101\relax
\mciteBstWouldAddEndPuncttrue
\mciteSetBstMidEndSepPunct{\mcitedefaultmidpunct}
{\mcitedefaultendpunct}{\mcitedefaultseppunct}\relax
\EndOfBibitem
\bibitem[Shea and Neuscamman(2017)Shea, and Neuscamman]{shea2017size}
Shea,~J.~A.; Neuscamman,~E. Size consistent excited states via algorithmic transformations between variational principles. \emph{Journal of Chemical Theory and Computation} \textbf{2017}, \emph{13}, 6078--6088\relax
\mciteBstWouldAddEndPuncttrue
\mciteSetBstMidEndSepPunct{\mcitedefaultmidpunct}
{\mcitedefaultendpunct}{\mcitedefaultseppunct}\relax
\EndOfBibitem
\bibitem[Garner and Neuscamman(2020)Garner, and Neuscamman]{garner2020variational}
Garner,~S.~M.; Neuscamman,~E. A variational Monte Carlo approach for core excitations. \emph{The Journal of chemical physics} \textbf{2020}, \emph{153}, 144108\relax
\mciteBstWouldAddEndPuncttrue
\mciteSetBstMidEndSepPunct{\mcitedefaultmidpunct}
{\mcitedefaultendpunct}{\mcitedefaultseppunct}\relax
\EndOfBibitem
\bibitem[Otis \latin{et~al.}(2020)Otis, Craig, and Neuscamman]{otis2020hybrid}
Otis,~L.; Craig,~I.~M.; Neuscamman,~E. A hybrid approach to excited-state-specific variational Monte Carlo and doubly excited states. \emph{The Journal of Chemical Physics} \textbf{2020}, \emph{153}, 234105\relax
\mciteBstWouldAddEndPuncttrue
\mciteSetBstMidEndSepPunct{\mcitedefaultmidpunct}
{\mcitedefaultendpunct}{\mcitedefaultseppunct}\relax
\EndOfBibitem
\bibitem[Shepard \latin{et~al.}(2022)Shepard, Panad{\'e}s-Barrueta, Moroni, Scemama, and Filippi]{shepard2022double}
Shepard,~S.; Panad{\'e}s-Barrueta,~R.~L.; Moroni,~S.; Scemama,~A.; Filippi,~C. Double excitation energies from quantum Monte Carlo using state-specific energy optimization. \emph{Journal of chemical theory and computation} \textbf{2022}, \emph{18}, 6722--6731\relax
\mciteBstWouldAddEndPuncttrue
\mciteSetBstMidEndSepPunct{\mcitedefaultmidpunct}
{\mcitedefaultendpunct}{\mcitedefaultseppunct}\relax
\EndOfBibitem
\bibitem[Otis and Neuscamman(2023)Otis, and Neuscamman]{otis2023optimization}
Otis,~L.; Neuscamman,~E. Optimization stability in excited-state-specific Variational Monte Carlo. \emph{Journal of Chemical Theory and Computation} \textbf{2023}, \emph{19}, 767--782\relax
\mciteBstWouldAddEndPuncttrue
\mciteSetBstMidEndSepPunct{\mcitedefaultmidpunct}
{\mcitedefaultendpunct}{\mcitedefaultseppunct}\relax
\EndOfBibitem
\bibitem[Otis and Neuscamman(2023)Otis, and Neuscamman]{otis2023promising}
Otis,~L.; Neuscamman,~E. A promising intersection of excited-state-specific methods from quantum chemistry and quantum Monte Carlo. \emph{Wiley Interdisciplinary Reviews: Computational Molecular Science} \textbf{2023}, \emph{13}, e1659\relax
\mciteBstWouldAddEndPuncttrue
\mciteSetBstMidEndSepPunct{\mcitedefaultmidpunct}
{\mcitedefaultendpunct}{\mcitedefaultseppunct}\relax
\EndOfBibitem
\bibitem[Pathak \latin{et~al.}(2021)Pathak, Busemeyer, Rodrigues, and Wagner]{pathak2021excited}
Pathak,~S.; Busemeyer,~B.; Rodrigues,~J.~N.; Wagner,~L.~K. Excited states in variational Monte Carlo using a penalty method. \emph{The Journal of Chemical Physics} \textbf{2021}, \emph{154}, 034101\relax
\mciteBstWouldAddEndPuncttrue
\mciteSetBstMidEndSepPunct{\mcitedefaultmidpunct}
{\mcitedefaultendpunct}{\mcitedefaultseppunct}\relax
\EndOfBibitem
\bibitem[Entwistle \latin{et~al.}(2023)Entwistle, Sch{\"a}tzle, Erdman, Hermann, and No{\'e}]{entwistle2023electronic}
Entwistle,~M.~T.; Sch{\"a}tzle,~Z.; Erdman,~P.~A.; Hermann,~J.; No{\'e},~F. Electronic excited states in deep variational Monte Carlo. \emph{Nature Communications} \textbf{2023}, \emph{14}, 274\relax
\mciteBstWouldAddEndPuncttrue
\mciteSetBstMidEndSepPunct{\mcitedefaultmidpunct}
{\mcitedefaultendpunct}{\mcitedefaultseppunct}\relax
\EndOfBibitem
\bibitem[Hanscam and Neuscamman(2022)Hanscam, and Neuscamman]{hanscam2022applying}
Hanscam,~R.; Neuscamman,~E. Applying generalized variational principles to excited-state-specific complete active space self-consistent field theory. \emph{Journal of Chemical Theory and Computation} \textbf{2022}, \emph{18}, 6608--6621\relax
\mciteBstWouldAddEndPuncttrue
\mciteSetBstMidEndSepPunct{\mcitedefaultmidpunct}
{\mcitedefaultendpunct}{\mcitedefaultseppunct}\relax
\EndOfBibitem
\bibitem[Roos \latin{et~al.}(1992)Roos, Andersson, and Fülscher]{roos_towards_1992}
Roos,~B.~O.; Andersson,~K.; Fülscher,~M.~P. Towards an accurate molecular orbital theory for excited states: the benzene molecule. \emph{Chemical Physics Letters} \textbf{1992}, \emph{192}, 5--13\relax
\mciteBstWouldAddEndPuncttrue
\mciteSetBstMidEndSepPunct{\mcitedefaultmidpunct}
{\mcitedefaultendpunct}{\mcitedefaultseppunct}\relax
\EndOfBibitem
\bibitem[Boyn and Mazziotti(2022)Boyn, and Mazziotti]{boyn_elucidating_2022}
Boyn,~J.-N.; Mazziotti,~D.~A. Elucidating the molecular orbital dependence of the total electronic energy in multireference problems. \emph{The Journal of Chemical Physics} \textbf{2022}, \emph{156}, 194104\relax
\mciteBstWouldAddEndPuncttrue
\mciteSetBstMidEndSepPunct{\mcitedefaultmidpunct}
{\mcitedefaultendpunct}{\mcitedefaultseppunct}\relax
\EndOfBibitem
\bibitem[Clune \latin{et~al.}(2020)Clune, Shea, and Neuscamman]{clune2020n5}
Clune,~R.; Shea,~J.~A.; Neuscamman,~E. N5-scaling excited-state-specific perturbation theory. \emph{Journal of chemical theory and computation} \textbf{2020}, \emph{16}, 6132--6141\relax
\mciteBstWouldAddEndPuncttrue
\mciteSetBstMidEndSepPunct{\mcitedefaultmidpunct}
{\mcitedefaultendpunct}{\mcitedefaultseppunct}\relax
\EndOfBibitem
\bibitem[Clune \latin{et~al.}(2023)Clune, Shea, Hardikar, Tuckman, and Neuscamman]{clune2023studying}
Clune,~R.; Shea,~J.~A.; Hardikar,~T.~S.; Tuckman,~H.; Neuscamman,~E. Studying excited-state-specific perturbation theory on the Thiel set. \emph{The Journal of Chemical Physics} \textbf{2023}, \emph{158}\relax
\mciteBstWouldAddEndPuncttrue
\mciteSetBstMidEndSepPunct{\mcitedefaultmidpunct}
{\mcitedefaultendpunct}{\mcitedefaultseppunct}\relax
\EndOfBibitem
\bibitem[Mayhall and Raghavachari(2010)Mayhall, and Raghavachari]{mayhall2010multiple}
Mayhall,~N.~J.; Raghavachari,~K. Multiple solutions to the single-reference CCSD equations for NiH. \emph{Journal of Chemical Theory and Computation} \textbf{2010}, \emph{6}, 2714--2720\relax
\mciteBstWouldAddEndPuncttrue
\mciteSetBstMidEndSepPunct{\mcitedefaultmidpunct}
{\mcitedefaultendpunct}{\mcitedefaultseppunct}\relax
\EndOfBibitem
\bibitem[Zheng and Cheng(2019)Zheng, and Cheng]{zheng2019performance}
Zheng,~X.; Cheng,~L. Performance of delta-coupled-cluster methods for calculations of core-ionization energies of first-row elements. \emph{Journal of chemical theory and computation} \textbf{2019}, \emph{15}, 4945--4955\relax
\mciteBstWouldAddEndPuncttrue
\mciteSetBstMidEndSepPunct{\mcitedefaultmidpunct}
{\mcitedefaultendpunct}{\mcitedefaultseppunct}\relax
\EndOfBibitem
\bibitem[Lee \latin{et~al.}(2019)Lee, Small, and Head-Gordon]{lee2019excited}
Lee,~J.; Small,~D.~W.; Head-Gordon,~M. Excited states via coupled cluster theory without equation-of-motion methods: Seeking higher roots with application to doubly excited states and double core hole states. \emph{The Journal of chemical physics} \textbf{2019}, \emph{151}, 214103\relax
\mciteBstWouldAddEndPuncttrue
\mciteSetBstMidEndSepPunct{\mcitedefaultmidpunct}
{\mcitedefaultendpunct}{\mcitedefaultseppunct}\relax
\EndOfBibitem
\bibitem[Damour \latin{et~al.}(2024)Damour, Scemama, Jacquemin, Kossoski, and Loos]{damour2024state}
Damour,~Y.; Scemama,~A.; Jacquemin,~D.; Kossoski,~F.; Loos,~P.-F. State-specific coupled-cluster methods for excited states. \emph{Journal of Chemical Theory and Computation} \textbf{2024}, \relax
\mciteBstWouldAddEndPunctfalse
\mciteSetBstMidEndSepPunct{\mcitedefaultmidpunct}
{}{\mcitedefaultseppunct}\relax
\EndOfBibitem
\bibitem[Kossoski \latin{et~al.}(2021)Kossoski, Marie, Scemama, Caffarel, and Loos]{kossoski2021excited}
Kossoski,~F.; Marie,~A.; Scemama,~A.; Caffarel,~M.; Loos,~P.-F. Excited States from State-Specific Orbital-Optimized Pair Coupled Cluster. \emph{Journal of Chemical Theory and Computation} \textbf{2021}, \emph{17}, 4756--4768\relax
\mciteBstWouldAddEndPuncttrue
\mciteSetBstMidEndSepPunct{\mcitedefaultmidpunct}
{\mcitedefaultendpunct}{\mcitedefaultseppunct}\relax
\EndOfBibitem
\bibitem[Tuckman and Neuscamman(2023)Tuckman, and Neuscamman]{tuckman_excited-state-specific_2023}
Tuckman,~H.; Neuscamman,~E. Excited-{State}-{Specific} {Pseudoprojected} {Coupled}-{Cluster} {Theory}. \emph{Journal of Chemical Theory and Computation} \textbf{2023}, \emph{19}, 6160--6171, Publisher: American Chemical Society\relax
\mciteBstWouldAddEndPuncttrue
\mciteSetBstMidEndSepPunct{\mcitedefaultmidpunct}
{\mcitedefaultendpunct}{\mcitedefaultseppunct}\relax
\EndOfBibitem
\bibitem[Liu \latin{et~al.}(2012)Liu, Fatehi, Shao, Veldkamp, and Subotnik]{liu2012communication}
Liu,~X.; Fatehi,~S.; Shao,~Y.; Veldkamp,~B.~S.; Subotnik,~J.~E. Communication: Adjusting charge transfer state energies for configuration interaction singles: Without any parameterization and with minimal cost. \emph{The Journal of chemical physics} \textbf{2012}, \emph{136}, 161101\relax
\mciteBstWouldAddEndPuncttrue
\mciteSetBstMidEndSepPunct{\mcitedefaultmidpunct}
{\mcitedefaultendpunct}{\mcitedefaultseppunct}\relax
\EndOfBibitem
\bibitem[Kossoski and Loos(2023)Kossoski, and Loos]{kossoski2022state}
Kossoski,~F.; Loos,~P.-F. State-specific configuration interaction for excited states. \emph{Journal of Chemical Theory and Computation} \textbf{2023}, \emph{19}, 2258--2269\relax
\mciteBstWouldAddEndPuncttrue
\mciteSetBstMidEndSepPunct{\mcitedefaultmidpunct}
{\mcitedefaultendpunct}{\mcitedefaultseppunct}\relax
\EndOfBibitem
\bibitem[Kossoski and Loos(2023)Kossoski, and Loos]{kossoski2023seniority}
Kossoski,~F.; Loos,~P.-F. Seniority and hierarchy configuration interaction for radicals and excited states. \emph{Journal of Chemical Theory and Computation} \textbf{2023}, \emph{19}, 8654--8670\relax
\mciteBstWouldAddEndPuncttrue
\mciteSetBstMidEndSepPunct{\mcitedefaultmidpunct}
{\mcitedefaultendpunct}{\mcitedefaultseppunct}\relax
\EndOfBibitem
\bibitem[Burton(2022)]{burton2022energy}
Burton,~H.~G. Energy Landscape of State-Specific Electronic Structure Theory. \emph{Journal of Chemical Theory and Computation} \textbf{2022}, \emph{18}, 1512--1526\relax
\mciteBstWouldAddEndPuncttrue
\mciteSetBstMidEndSepPunct{\mcitedefaultmidpunct}
{\mcitedefaultendpunct}{\mcitedefaultseppunct}\relax
\EndOfBibitem
\bibitem[Tsuchimochi(2024)]{tsuchimochi2024CISthenCIS}
Tsuchimochi,~T. Double configuration interaction singles: Scalable and size-intensive approach for orbital relaxation in excited states and bond-dissociation. \emph{The Journal of Chemical Physics} \textbf{2024}, \emph{161}\relax
\mciteBstWouldAddEndPuncttrue
\mciteSetBstMidEndSepPunct{\mcitedefaultmidpunct}
{\mcitedefaultendpunct}{\mcitedefaultseppunct}\relax
\EndOfBibitem
\bibitem[Loos \latin{et~al.}(2018)Loos, Scemama, Blondel, Garniron, Caffarel, and Jacquemin]{loos_mountaineering_2018}
Loos,~P.-F.; Scemama,~A.; Blondel,~A.; Garniron,~Y.; Caffarel,~M.; Jacquemin,~D. A {Mountaineering} {Strategy} to {Excited} {States}: {Highly} {Accurate} {Reference} {Energies} and {Benchmarks}. \emph{Journal of Chemical Theory and Computation} \textbf{2018}, \emph{14}, 4360--4379, Publisher: American Chemical Society\relax
\mciteBstWouldAddEndPuncttrue
\mciteSetBstMidEndSepPunct{\mcitedefaultmidpunct}
{\mcitedefaultendpunct}{\mcitedefaultseppunct}\relax
\EndOfBibitem
\bibitem[Loos \latin{et~al.}(2020)Loos, Lipparini, Boggio-Pasqua, Scemama, and Jacquemin]{loos_mountaineering_2020}
Loos,~P.-F.; Lipparini,~F.; Boggio-Pasqua,~M.; Scemama,~A.; Jacquemin,~D. A {Mountaineering} {Strategy} to {Excited} {States}: {Highly} {Accurate} {Energies} and {Benchmarks} for {Medium} {Sized} {Molecules}. \emph{Journal of Chemical Theory and Computation} \textbf{2020}, \emph{16}, 1711--1741, Publisher: American Chemical Society\relax
\mciteBstWouldAddEndPuncttrue
\mciteSetBstMidEndSepPunct{\mcitedefaultmidpunct}
{\mcitedefaultendpunct}{\mcitedefaultseppunct}\relax
\EndOfBibitem
\bibitem[Martin(2003)]{martin2003natural}
Martin,~R.~L. Natural transition orbitals. \emph{The Journal of chemical physics} \textbf{2003}, \emph{118}, 4775--4777\relax
\mciteBstWouldAddEndPuncttrue
\mciteSetBstMidEndSepPunct{\mcitedefaultmidpunct}
{\mcitedefaultendpunct}{\mcitedefaultseppunct}\relax
\EndOfBibitem
\bibitem[Eckart and Young(1936)Eckart, and Young]{eckart1936approximation}
Eckart,~C.; Young,~G. The approximation of one matrix by another of lower rank. \emph{Psychometrika} \textbf{1936}, \emph{1}, 211--218\relax
\mciteBstWouldAddEndPuncttrue
\mciteSetBstMidEndSepPunct{\mcitedefaultmidpunct}
{\mcitedefaultendpunct}{\mcitedefaultseppunct}\relax
\EndOfBibitem
\bibitem[Furche(2001)]{furche2001density}
Furche,~F. On the density matrix based approach to time-dependent density functional response theory. \emph{The Journal of Chemical Physics} \textbf{2001}, \emph{114}, 5982--5992\relax
\mciteBstWouldAddEndPuncttrue
\mciteSetBstMidEndSepPunct{\mcitedefaultmidpunct}
{\mcitedefaultendpunct}{\mcitedefaultseppunct}\relax
\EndOfBibitem
\bibitem[Sun \latin{et~al.}(2020)Sun, Zhang, Banerjee, Bao, Barbry, Blunt, Bogdanov, Booth, Chen, Cui, \latin{et~al.} others]{sun2020recent}
Sun,~Q.; Zhang,~X.; Banerjee,~S.; Bao,~P.; Barbry,~M.; Blunt,~N.~S.; Bogdanov,~N.~A.; Booth,~G.~H.; Chen,~J.; Cui,~Z.-H.; others Recent developments in the PySCF program package. \emph{The Journal of chemical physics} \textbf{2020}, \emph{153}, 024109\relax
\mciteBstWouldAddEndPuncttrue
\mciteSetBstMidEndSepPunct{\mcitedefaultmidpunct}
{\mcitedefaultendpunct}{\mcitedefaultseppunct}\relax
\EndOfBibitem
\bibitem[Mardirossian and Head-Gordon(2014)Mardirossian, and Head-Gordon]{mardirossian2014omegab97x}
Mardirossian,~N.; Head-Gordon,~M. $\omega$B97X-V: A 10-parameter, range-separated hybrid, generalized gradient approximation density functional with nonlocal correlation, designed by a survival-of-the-fittest strategy. \emph{Physical Chemistry Chemical Physics} \textbf{2014}, \emph{16}, 9904--9924\relax
\mciteBstWouldAddEndPuncttrue
\mciteSetBstMidEndSepPunct{\mcitedefaultmidpunct}
{\mcitedefaultendpunct}{\mcitedefaultseppunct}\relax
\EndOfBibitem
\bibitem[Epifanovsky \latin{et~al.}(2021)Epifanovsky, Gilbert, Feng, Lee, Mao, Mardirossian, Pokhilko, White, Coons, Dempwolff, \latin{et~al.} others]{epifanovsky2021software}
Epifanovsky,~E.; Gilbert,~A.~T.; Feng,~X.; Lee,~J.; Mao,~Y.; Mardirossian,~N.; Pokhilko,~P.; White,~A.~F.; Coons,~M.~P.; Dempwolff,~A.~L.; others Software for the frontiers of quantum chemistry: An overview of developments in the Q-Chem 5 package. \emph{The Journal of chemical physics} \textbf{2021}, \emph{155}\relax
\mciteBstWouldAddEndPuncttrue
\mciteSetBstMidEndSepPunct{\mcitedefaultmidpunct}
{\mcitedefaultendpunct}{\mcitedefaultseppunct}\relax
\EndOfBibitem
\bibitem[Laurent and Jacquemin(2013)Laurent, and Jacquemin]{laurent2013td}
Laurent,~A.~D.; Jacquemin,~D. TD-DFT benchmarks: a review. \emph{International Journal of Quantum Chemistry} \textbf{2013}, \emph{113}, 2019--2039\relax
\mciteBstWouldAddEndPuncttrue
\mciteSetBstMidEndSepPunct{\mcitedefaultmidpunct}
{\mcitedefaultendpunct}{\mcitedefaultseppunct}\relax
\EndOfBibitem
\end{mcitethebibliography}

\clearpage
\onecolumngrid
\section{Supplementary Information}
\renewcommand{\thesection}{S\arabic{section}}
\renewcommand{\theequation}{S\arabic{equation}}
\renewcommand{\thefigure}{S\arabic{figure}}
\renewcommand{\thetable}{S\arabic{table}}
\setcounter{section}{0}
\setcounter{figure}{0}
\setcounter{equation}{0}
\setcounter{table}{0}
\subsection{Multi-CSF and State Inclusion Details} 
\FloatBarrier
\begin{figure*}[]
    \centering
    \includegraphics[width=0.95\linewidth]{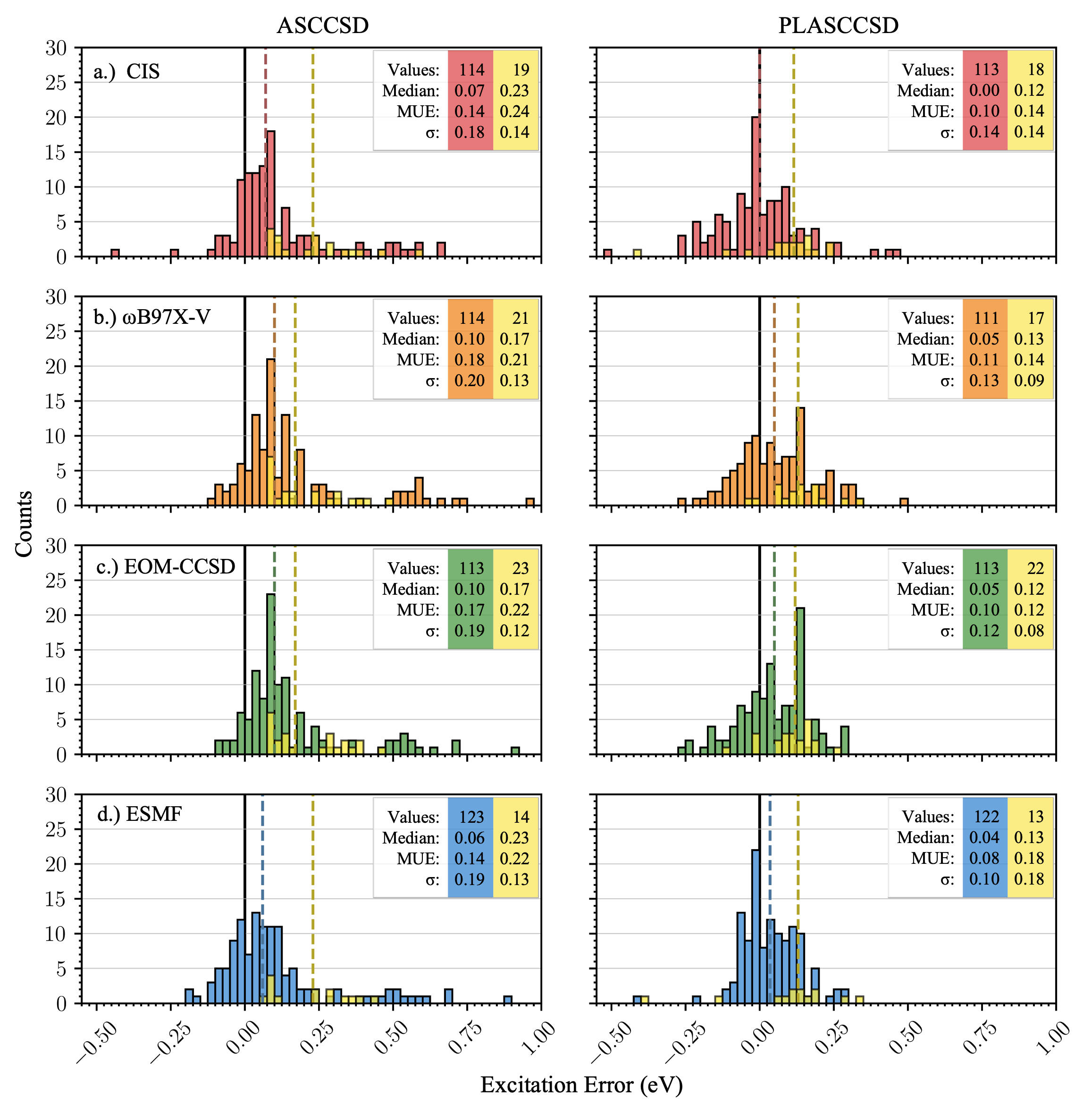}
    \caption{ASCC (left) and PLASCC (right) excitation energy error distributions on the QUEST valence and Rydberg states
    for different starting points broken out by one-CSF (shown in color) or two-CSF (shown in yellow).
    Dashed lines show median errors, while these along with the MUEs,
    standard deviations, and number of converged states
    are shown in the insets.}
    \label{fig:small_med_nnto}
\end{figure*}

The determination of the wave function being single
or multi-CSF is starting point dependent.
Of the 138 valence and Rydberg states 14 were considered two-CSF by ESMF, 23 by CIS, 22 by $\omega$B97X-V, and 
23 by EOM-CCSD. In Fig.~\ref{fig:small_med_nnto} we breakdown the results of ASCC and PLASCC on the QUEST set of valence and Rydberg states by one-CSF, shown in color, or two-CSF shown in yellow. This delineation between the different CSF type wave functions reveals that the two-CSF states consistently overestimate the excitation energies relative to the one-CSF states. When compared to ASCC, PLASCC slightly improves on the two-CSF states, reducing the median relative to ASCC anywhere from 0.04 to 0.11 eV.  
%
Moving on to the specifics of the winnowing procedure among our reduced valence and Rydberg test set,
five individual cases were removed for being two-CSF with more than four non-hydrogen atoms. 
These include four CIS states --- thiophene $1^1B_1$, 
tetrazine $1^1B_{2g}$, 
pyrazine $2^1B_{1g}$, 
and pyrimidine $2^1B_1$ ---
and one $\omega$B97X-V state --- thiophene $2^1B_2$.
The $\omega$B97X-V reference for pyridazine $1^1B_2$ was omitted due to being a three-CSF.
Six states 
across ASCCSD and PLASCCSD were not reported due to convergence issues, specifically ammonia $2^1 A_2$ using an EOM-CCSD reference for ASCC and 
five $\omega$B97X-V states ---
formaldehyde $1^1B_2$,  
formamide $3^1A'$,      
isobutene $1^1B_1$,
thioacetone $2^1A_1$,   
and cyclopropenone $3^1B_2$ ---
for PLASCC.    
Lastly in the Quest set of result, all four references were removed for furan $2^1B_2$ in ASCC and
cyclopropenethione $1^1B_2$ and $1^1B_2$
as well as diacetylene $1^1\Delta_u$ in PLASCC due to not having results in at least two of the references considered. 
In the charge transfer set, PLASCC results on two CIS states ---
ammonia-oxygendifluoride $4^1A’$ and
3,5-difluoro-penta-2,4-dienamine $1^1A^"$ --- are left unreported due to convergence issues.

\subsection{Raw Data}
The table below includes excitation energies (eV) of EMSF, CIS, TD-DFT/$\omega$B97X-V, and EOM-CCSD along with their corresponding ASCCSD and PLASCCSD counterparts.
The 'Flip' column indicates states with two similar but distinct ansatz choices for the ASCC framework.~\cite{tuckman_improving_2025} Both energies are reported.
The 'Rydberg' column indicates Rydberg states plotted separately in Fig.~\ref{fig:small_med_rydberg}, as designated by the QUEST small and medium sets.~\
The '\textit{n-CSF}' column indicates how many singular values were >0.2 when forming the reference for each of the four methods (only \textit{n}= 1 or 2 were calculated).
The 'REF' column are reference energies.
Reference energies for the QUEST small and medium sets are at exFCI for molecules with three or fewer non-hydrogen
atoms and EOM-CCSDT otherwise unless explicitly stated.
Reference energies for the charge transfer set are at EOM-CCSDT unless explicitly stated. \\

\noindent Table annotations: *States are loosely converged, but energetically change below precision reported iteration by iteration.
(a) EOM-CCSDTQ reference. 
(b) EOM-CCSDT reference. 
(c) LR-CC3 reference. \\

\includegraphics[page=1]{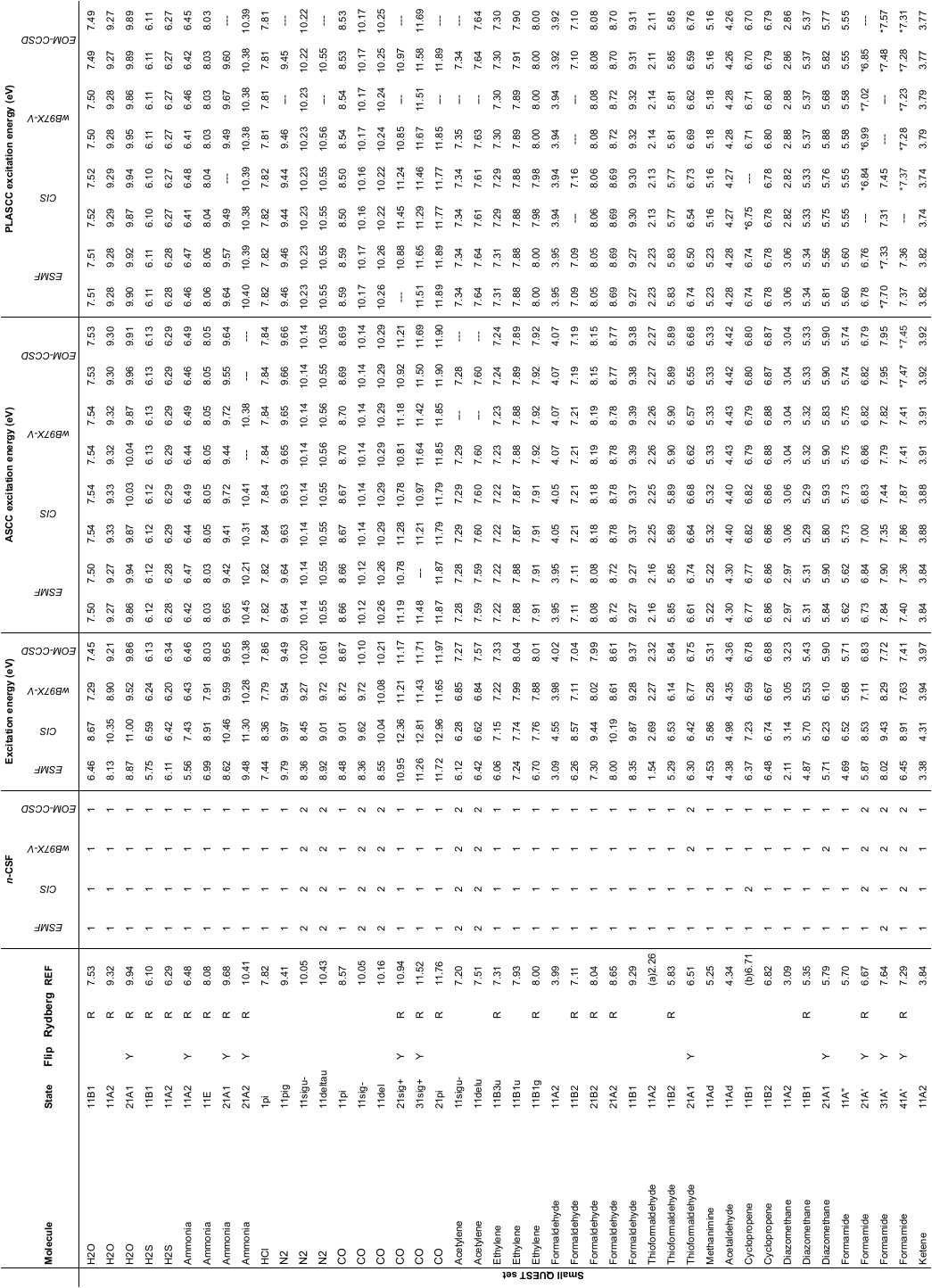}
\includegraphics[page=2]{SI_allResults.pdf}
\includegraphics[page=3]{SI_allResults.pdf}
\includegraphics[page=4]{SI_allResults.pdf}

\end{document}